\documentclass[aps,prd,twocolumn,nofootinbib,superscriptaddress]{revtex4-2}

\usepackage[T1]{fontenc}
\usepackage{lmodern}
\usepackage{amsmath,amssymb,bm}
\usepackage{graphicx}
\usepackage{hyperref}
\usepackage{siunitx}
\usepackage{xcolor}

\begin{document}

\title{Continuous coherent spin-frequency metrology in storage rings via resonant beam-driven detection}

\author{Younggeun Kim}
\affiliation{Johannes Gutenberg-Universit{\"a}t Mainz, 55122 Mainz, Germany}
\affiliation{Helmholtz Institute Mainz, 55099 Mainz, Germany}
\affiliation{GSI Helmholtzzentrum f\"ur Schwerionenforschung GmbH, 64291 Darmstadt, Germany}

\author{Themis Bowcock}
\affiliation{University of Liverpool, Liverpool, United Kingdom}

\author{Dmitry Budker}
\affiliation{Johannes Gutenberg-Universit{\"a}t Mainz, 55122 Mainz, Germany}
\affiliation{Helmholtz Institute Mainz, 55099 Mainz, Germany}
\affiliation{GSI Helmholtzzentrum f\"ur Schwerionenforschung GmbH, 64291 Darmstadt, Germany}
\affiliation{Department of Physics, University of California, Berkeley, CA 94720-7300, United States of America}

\author{Giovanni Cantatore}
\affiliation{
  University of Trieste and National Institute for Nuclear Physics (INFN-Trieste), Trieste, Italy}
\author{Hooman Davoudiasl}
\affiliation{Brookhaven National Laboratory, Upton, New York 11973, USA}

\author{Dmitry Denisov}
\affiliation{Brookhaven National Laboratory, Upton, New York 11973, USA}

\author{Abhay Deshpande}
\affiliation{Brookhaven National Laboratory, Upton, New York 11973, USA}

\author{Wolfram Fischer}
\affiliation{Brookhaven National Laboratory, Upton, New York 11973, USA}

\author{Selcuk Haciomeroglu}
\affiliation{Istinye University, Istanbul, T\"urkiye}

\author{Haixin Huang}
\affiliation{Brookhaven National Laboratory, Upton, New York 11973, USA}

\author{David Kawall}
\affiliation{Department of Physics, University of Massachusetts Amherst, Amherst, Massachusetts 01003, USA}

\author{Alexander Keshavarzi}
\affiliation{Department of Physics and Astronomy, University College London, London WC1E 6BT, United Kingdom}

\author{On Kim}
\affiliation{University of Washington, Seattle, Washington 98195, USA}

\author{Ivan Koop}
\affiliation{Budker Institute of Nuclear Physics, Novosibirsk, Russia}
  
\author{Valeri Lebedev}
\affiliation{Joint Institute for Nuclear Research, Dubna, Russia}

\author{Jonathan Lee}
\affiliation{Brookhaven National Laboratory, Upton, New York 11973, USA}

\author{William M. Morse}
\affiliation{Brookhaven National Laboratory, Upton, New York 11973, USA}

\author{Cenap Ozben}
\affiliation{
  Istanbul Technical University, Istanbul, T\"urkiye}

\author{Vincent Schoefer}
\affiliation{Brookhaven National Laboratory, Upton, New York 11973, USA}

\author{Yannis K. Semertzidis}
\email{semertzidisy@gmail.com}
\affiliation{Innovative Solutions R\&D LLC, Stony Brook, New York 11790, USA}

\author{Eleftherios Skordis}
\affiliation{Brookhaven National Laboratory, Upton, New York 11973, USA}

\author{Edward Stephenson}
\affiliation{Indiana University, Bloomington, Indiana 47405, USA}

\author{Vladimir Tishchenko}
\affiliation{Brookhaven National Laboratory, Upton, New York 11973, USA}

\author{Nicholaos Tsoupas}
\affiliation{Brookhaven National Laboratory, Upton, New York 11973, USA}

\author{Graziano Venanzoni}
\affiliation{University of Liverpool, Liverpool, United Kingdom}
\affiliation{INFN, Sezione di Pisa, Pisa, Italy}

\author{Joost Vossebeld}
\affiliation{University of Liverpool, Liverpool, United Kingdom}

\author{Peter Winter}
\affiliation{Argonne National Laboratory, Argonne, Illinois 60439, USA}

\date{\today}

\begin{abstract}
Precision measurements in storage rings are increasingly limited not by beam
intensity, but by the ability to monitor collective spin dynamics coherently
over long time scales.
Existing polarimetry techniques rely on destructive scattering processes that
preclude continuous, non-intercepting tracking of spin evolution, sample only a
small fraction of the stored beam, and fundamentally constrain both statistical
sensitivity and systematic control.

Here we introduce a non-destructive, phase-coherent polarimetry paradigm in
which the stored beam polarization is treated as a continuous dynamical
observable rather than a quantity inferred from rare scattering events.
Spin-dependent electromagnetic fields generated by a polarized relativistic
beam establish a symmetry-selected differential boundary condition on pickup
electrodes.
This signal is transduced into a narrowband phase modulation of a high-$Q$
resonator that is continuously interrogated with a coherent probe.
The resonator provides the essential narrowband filtering, while the
charge-induced pickup—dominant at the raw field level—is rejected through a
combination of geometric symmetry, helicity reversal, and synchronous
demodulation.

By enabling truly non-intercepting polarimetry, this architecture allows
storage-ring experiments to exploit coherence-preserving measurement strategies
developed in other precision fields, including atomic clocks
\cite{LudlowRevModPhys2015} and nuclear magnetic resonance
\cite{Kimball2016_PhysRevLett.116.190801,Ni2025_1v1p-kpb2}. 
In particular, controlled spin precession (spin-wheel operation) provides a
stable phase reference that shifts the measurement away from DC and enables
phase-coherent detection of slow spin evolution.
When combined with optimized lattice design and beam cooling, this approach
can substantially extend the usable spin coherence time, with values
approaching $10^{5}\,\mathrm{s}$ appearing realistic within existing
accelerator technology.

The resulting phase-coherent readout supports optimal slope-based estimation
with $T^{-3/2}$ statistical scaling~\cite{Kay1993,Schmitt2017T32}, while
eliminating the efficiency penalties inherent to scattering-based polarimetry.
For storage-ring EDM experiments, this combination enables sensitivity
approaching the level expected within the Standard Model without destructive
spin readout or the need for quantum-enhanced spin correlations.
More broadly, the method establishes a general phase-coherent measurement
architecture for collective spin dynamics in storage rings, directly adapting
resonant sensing concepts developed in axion dark-matter searches to
charged-particle precision experiments.
\end{abstract}

\maketitle

\section{Introduction}

Precision measurements of spin dynamics in storage rings provide a powerful
window onto physics beyond the Standard Model.
Searches for permanent electric dipole moments (EDMs), tests of discrete
symmetries, and probes of ultralight dark matter all rely on the ability to track
extremely small, coherent changes in spin polarization over extended periods of
time.
In many such experiments, the signal of interest is not a large instantaneous
effect, but a minute phase-coherent evolution that accumulates slowly over
hours or even days.
As experimental control of beams and spin manipulation has steadily improved,
it has become increasingly clear that a central limitation now lies elsewhere:
polarimetry.

Conventional storage-ring polarimetry relies on scattering from thin internal
targets, inferring polarization from asymmetries in rare beam--target
interactions.
While robust and well established, this approach is intrinsically intercepting.
Only a small fraction of the stored beam is sampled, and the observable is
suppressed by both the finite detection efficiency and the analyzing power of
the scattering process.
For proton scattering at proton kinetic energy
$T\simeq 233$~MeV (near the proton magic momentum~\cite{Omarov_2022_symmetric,Farley_PhysRevLett.93.052001}),
asymmetry values as large as $A\sim0.6$ can be achieved~\cite{Jeong:2022}, but at other energies the
analyzing power is typically significantly smaller.
With representative efficiencies $\varepsilon\sim1\%$, the effective
statistical weight is reduced by a factor $\varepsilon A^{2}\lesssim
\mathcal{O}(10^{-3})$.
These penalties severely limit continuous tracking of spin dynamics and make it
difficult to exploit long spin coherence times.

A qualitatively different paradigm is to treat the stored-beam polarization as a
continuous, phase-coherent dynamical variable and to monitor its evolution
non-destructively through the electromagnetic fields generated by the polarized
beam itself.
In this view, the experimental task is not to count discrete scattering events,
but to estimate the slope of a time-dependent phase whose evolution encodes the
underlying physics.
Sensitivity is therefore gained not by extracting power from the beam, but by
tracking a coherent phase over long durations.

In the present work we develop a coherent polarimetry framework based on
resonant electromagnetic pickup of the spin-dependent fields of a circulating
polarized beam.
A relativistic bunch carrying a collective magnetic moment generates a
spin-dependent magnetic field.
In the laboratory frame this produces a transverse electric field proportional
to $\mathbf{v}\times\mathbf{B}_{\rm spin}$.
When sampled by a pair of symmetric pickup electrodes, this field induces a
tiny differential voltage correlated with the beam polarization.
Because the beam is bunched~\cite{Omarov_2022_symmetric}, the pickup signal contains discrete Fourier
components at harmonics of the bunch frequency.
By coupling the electrodes to a high-quality ($Q$) narrowband LC resonator
tuned to one such harmonic (e.g.\ $18.18\,\mathrm{MHz}$ for the bunch pattern
considered here), the spin-dependent signal is coherently enhanced while
broadband backgrounds are rejected. The use of a specific harmonic of the
bunch structure follows the same general principle as harmonic-mode
manipulation employed in multi-harmonic RF kicker systems~\cite{Huang_harmonic_PhysRevAccelBeams.19.084201},
where controlled excitation of selected harmonics enables precise temporal
selectivity within a bunched beam.

A central challenge is the enormous dynamic range between the
charge-induced electromagnetic fields of the bunched beam and the much
smaller spin-dependent fields.
At the raw field level the charge response exceeds the spin-dependent
response by approximately $10^{14}$ for realistic storage-ring parameters.
Previous attempts to observe the spin-dependent fields of stored beams
have encountered this difficulty, including early theoretical proposals
and experimental studies based on resonant pickup
structures~\cite{cameron1997,cameron1996absolute,blaskiewicz1996absolute,derbenev1998radio},
as well as more recent work~\cite{Roberts2021Polarimeter}.
Rather than attempting to resolve this ratio directly, the present
experiment distributes the required suppression across the measurement
chain: symmetry rejection of the charge signal, resonant enhancement of
the spin signal by a high-$Q$ pickup resonator, and long coherent phase
integration enabled by the probing readout.

The charge signal is helicity-even and uncorrelated with controlled
spin manipulations, whereas the spin-dependent signal is helicity-odd and
phase-coherent.

Through symmetric pickup geometry, immediate left--right subtraction,
positive--negative helicity subtraction at the bunch level, resonant
selection at a chosen harmonic, and comparison of clockwise (CW) and
counter-clockwise (CCW) beams, the effective charge background is suppressed
by many orders of magnitude.
After these symmetry operations the residual charge contribution is reduced
relative to the spin signal and enters primarily as additive broadband noise,
so that an adequate SNR can be achieved at each sampling time (see
Table~\ref{tab:symmetry}).

The remaining challenge is the intrinsic smallness of the spin-induced
electromagnetic field itself.
This is addressed by operating the pickup in a high-$Q$ resonant mode that
selectively amplifies the coherent spin signal at the chosen harmonic.
The resonator is continuously interrogated with a coherent probing tone,
so that the spin-dependent field appears as a narrowband phase modulation of
the injected signal~\cite{ProbingPRD2023,CARAMEL_PRD2026}.
To first order the resonator output can be written as
\begin{equation}
V(t)=V_0\cos\!\left[\omega_p t+\delta\phi(t)\right],
\end{equation}
where $V_0$ and $\omega_p$ are the amplitude and frequency of the injected
probing signal and $\delta\phi(t)$ is the small phase perturbation generated
by the spin-dependent field.
This phase modulation produces narrow sidebands at
$\omega_p\pm\omega_{\rm SW}$, where $\omega_{\rm SW}$ is the applied
spin-modulation (``spin-wheel''~\cite{derbenev1998radio,Koop:2013vja,Koop:2015lez}) angular frequency. 

In the spin-wheel configuration, a deliberately applied
weak vertical field produces a controlled continuous spin
precession about the radial direction at a well-defined
frequency $\omega_{\rm SW}$. Rather than attempting to
measure a static vertical polarization buildup directly,
the EDM signal is thereby converted into a small shift of
a coherent oscillatory spin phase. The measurement is
thus moved away from DC into a narrowband frequency
channel, enabling phase-sensitive detection, improved
rejection of slow drifts and static systematic effects,
and direct application of coherent frequency-metrology
techniques.


This allows the signal to be extracted through phase-sensitive detection. Because the phase of the probe can be measured with extremely high precision, this heterodyne-like detection scheme converts a minute physical field into a measurable phase shift while rejecting broadband noise outside the resonant bandwidth.
Furthermore, since the modulation remains phase-coherent with the probing
tone, the signal can be integrated coherently over long times, providing an
additional effective gain that scales approximately as $1 / \sqrt{T}$ with the
integration time $T$, while the overall corresponding frequency sensitivity improves
as $T^{-3/2}$.

Operationally, the experiment does not measure the tiny
spin-dependent field amplitude directly. Instead, the
spin-dependent field produces a small phase modulation
of the resonantly enhanced carrier generated by the
combined beam harmonic and coherent probing tone. The
measured observable is therefore the time-dependent
demodulated phase $\phi(t)$ of the resonator output.
During spin-wheel operation this phase contains a
well-defined oscillatory component at the spin-wheel
frequency, while the EDM appears as a small differential
frequency shift between the CW and CCW phase
evolutions. The experiment therefore reduces to a
precision frequency (phase-slope) measurement rather
than to a direct measurement of a tiny instantaneous
electromagnetic field amplitude.

\begin{table*}[t]
\centering
\caption{Hierarchy of symmetry-based reversals and expected suppression of charge-dominated backgrounds, for justification see below in Methods~\ref{subsec:methods_charge_symmetry}. Orders of magnitude are conservative estimates for realistic storage-ring parameters.}
\begin{tabular}{lcc}
\hline
Symmetry / Reversal Mechanism & Physical Principle & Suppression (orders of magnitude) \\
\hline
Left--Right electrode subtraction & Geometric antisymmetry of charge field & $\sim 10^{3}$--$10^{4}$ \\
Positive--Negative helicity subtraction & Helicity-even charge vs helicity-odd spin & $\sim 10^{4}$--$10^{5}$ \\
Spin-wheel modulation (0.1--10 Hz) & Frequency separation from DC charge response &  $\sim 10^{3}$--$10^{4}$ \\
CW vs.\ CCW comparison & Opposite spin-wheel rotation; EDM common-mode &  $\sim 10^{3}$--$10^{4}$ \\
\hline
Total effective suppression & Combined symmetry hierarchy & $\sim 10^{13}$--$10^{17}$ \\
\hline
\end{tabular}
\label{tab:symmetry}
\end{table*}

The controlled vertical spin precession, or ``spin wheel~\cite{derbenev1998radio},'' first introduced
in the storage-ring EDM context by Koop~\cite{Koop:2013vja,Koop:2015lez},
provides a well-defined phase reference for synchronous detection.
In the present architecture the spin-wheel frequency is typically in the
$0.1$--$10$~Hz range. 
Importantly, the EDM-induced phase evolution is common-mode between CW and
CCW beams, whereas the magnetic spin-wheel rotation occurs in opposite
directions for the two beams.
This separation in symmetry space provides an additional powerful rejection
handle against systematic effects.

The coherent polarimeter does not alter the fundamental statistical scaling of
spin-based measurements.
For a linear-in-time signal extracted over a coherent interval $T$, the
uncertainty of an optimal slope estimator scales as $T^{-3/2}$~\cite{Kay1993}.
Its decisive advantage lies in enabling substantially longer coherent
integration by eliminating the efficiency and intensity penalties inherent to
scattering-based readout.
Because the measurement is non-intercepting, the full stored beam contributes
continuously, avoiding beam depletion and measurement-induced decoherence.

\begin{figure*}[t]
  \centering
  \includegraphics[width=\textwidth]{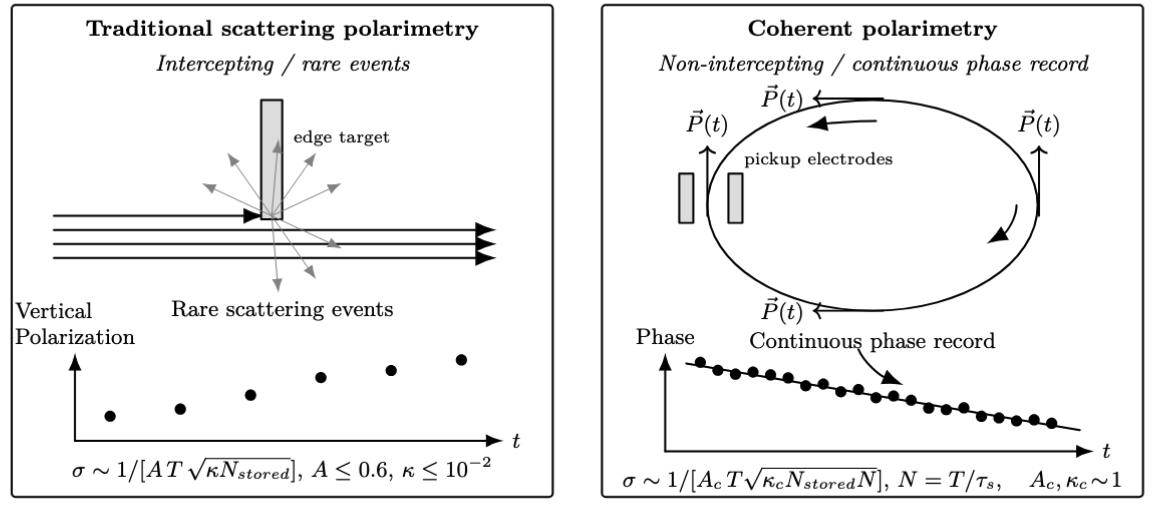}
  \caption{Conceptual comparison between conventional scattering polarimetry
  and coherent, non-destructive polarimetry.
  (\textit{Left}) In scattering-based polarimetry, only a small fraction of the
  stored beam is sampled, and the statistical sensitivity is reduced by the
  product of detection efficiency ($\kappa$) and analyzing power ($A$).
  (\textit{Right}) In the coherent polarimeter, the entire stored beam contributes
  continuously to a phase-coherent signal, eliminating efficiency and
  analyzing-power penalties and enabling long coherent integration. $T$ is the beam spin coherence time (SCT) and $\tau_s$ is the sample time, as defined by the detection system, see discussion in Methods.}
  \label{fig:comparison}
\end{figure*}

The applicability of the Cram\'er--Rao (CR) bound in this context requires that
the demodulated phase time series satisfy the usual regularity conditions~\cite{Kay1993}:
stationary noise, short-range correlations, and a sufficiently large
signal-to-noise ratio~\cite{RifeBoorstyn1974,Quinn1994} at each effective sampling interval.
In particular, the CR bound assumes that the per-sample phase estimate has
adequate instantaneous SNR so that the likelihood function is locally
quadratic and unbiased estimators can be constructed~\cite{Tretter1985,Boashash1992,Kay1993,Tretter1985}.
In the present framework, the narrowband resonant filtering and helicity-tagged
subtraction ensure that the baseband phase noise is dominated by approximately
Gaussian additive fluctuations, while the effective sampling interval is set by
the resonator bandwidth and downstream filtering.
Under these conditions least-squares slope fitting is statistically efficient
and saturates the CR bound, yielding the $T^{-3/2}$ scaling derived in
Methods.
The qualitative distinction between conventional scattering-based polarimetry
and coherent, non-destructive polarimetry is illustrated schematically in
Fig.~\ref{fig:comparison}. 

\begin{figure*}[t]
  \vspace{0cm}
  \centering
  \includegraphics[width=\textwidth]{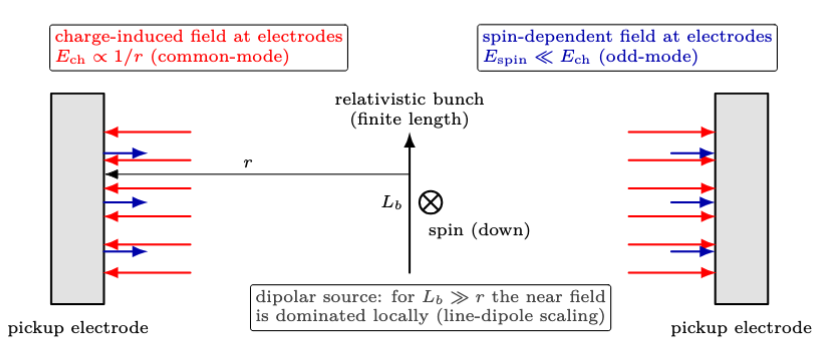}
  \vspace{0cm}
  \caption{
  Sketch of the field-scaling estimates used in
Eqs.~(\ref{eq:Espin_scaling})--(\ref{eq:Echarge_scaling}), drawn as the electric fields \emph{at the pickup electrodes}
in a top-view geometry between two parallel plates (not to scale).
The charge-induced near field (red) appears predominantly as a common-mode contribution at the opposing
electrode faces, with characteristic scaling $E_{\rm ch}\propto 1/r$ for a finite relativistic line charge.
The spin-dependent contribution (dark blue) is much smaller and is indicated here by arrows whose length is
drawn at $1/2$ of the charge-induced arrows for visual clarity.
The vertical polarization during the spin wheel is represented by an out-of-plane marker.
For the representative parameters considered here, the bunch length is $L_b\simeq 1~\mathrm{m}$,
while the characteristic transverse beam--electrode separation is $r\simeq 2~\mathrm{cm}$
(corresponding to a plate spacing $2r\simeq 4~\mathrm{cm}$); the electrodes integrate the resulting
boundary-condition perturbation over the full bunch passage, while the dipolar spin field differs from
the monopolar charge field only in its geometric scaling and overall magnitude.
}
\label{fig:Fields_on_PE}
\end{figure*}

\begin{figure*}[t]
  \centering
  \includegraphics[width=\textwidth]{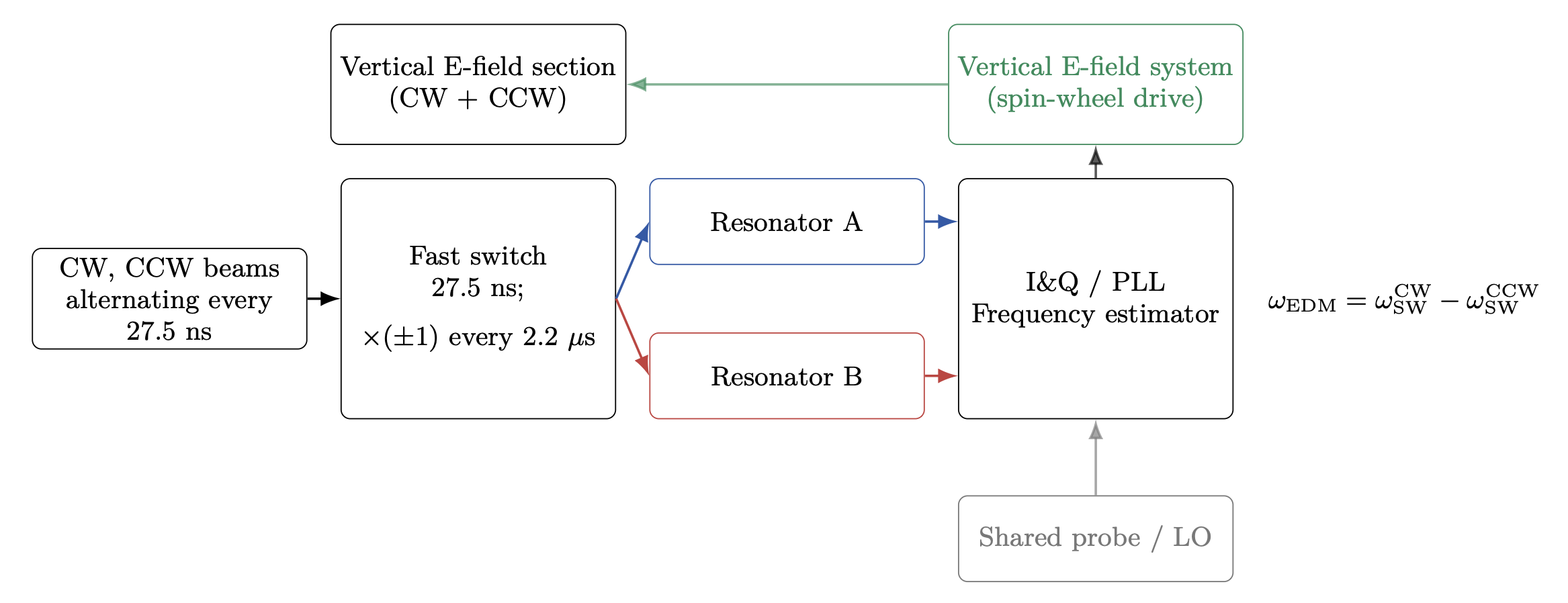}
  \caption{After left--right subtraction of the pickup-electrode (PE) signals to
suppress the dominant charge-induced response, the differential signal is
routed by a fast switch that demultiplexes the alternating clockwise (CW) and
counterclockwise (CCW) bunches and forwards them to their respective
resonators. Prior to injection, the signals are multiplied by $\pm 1$
according to their helicity tagging, further reducing residual charge pickup
while preserving the spin-dependent component. The PE plates operate at room
temperature, whereas the resonators are maintained at 4~K and are driven by a
common reference (probe) tone at 18.18~MHz. The spin-induced signal phase
modulates the resonator field, producing sidebands about the carrier. A
feedback loop stabilizes the spin-wheel frequency in one channel, and the
electric dipole moment (EDM) observable is encoded entirely in the difference
between the spin-wheel frequencies of the counter-rotating beams.
Longitudinal bunch-timing fluctuations and synchrotron
motion can produce substantial common-mode phase
modulation of the carrier at the selected bunch harmonic.
These effects are expected to enter primarily as
charge-dominated technical phase noise and are
suppressed by the helicity-tagged CW/CCW differential
measurement architecture.}
  \label{fig:Res_PLL}
\end{figure*}

For a proton bunch whose polarization is vertically downward, the magnetic
dipole moment of the beam is aligned with the vertical direction. The
collective magnetic field associated with this moving magnetization
produces, in the laboratory frame, a transverse electric field whose
magnitude scales as
\[
E_{\rm spin}\sim \beta c\,B_{\rm spin}.
\]
For the pickup-electrode geometry shown in Fig.~\ref{fig:Fields_on_PE},
this field has a radial component and induces opposite surface charges
on the left and right PE plates.
 The resulting electrode signals therefore have opposite sign,
providing the antisymmetric component that carries the spin-dependent
information.
The left--right signals from the pickup electrodes (PE) are combined in a
hybrid subtractor to suppress the dominant charge-induced response, thereby
isolating the spin-dependent component, as illustrated in
Fig.~\ref{fig:Res_PLL}. The resulting differential signal, corresponding to
alternating CW and CCW bunches separated by $27.5~\mathrm{ns}$, is routed by a
fast switch to the appropriate resonator channel. Prior to injection into the
resonators, the signal is multiplied by $\pm 1$ according to the helicity of
each bunch (period $2.2~\mu\mathrm{s}$), thereby implementing synchronous
helicity tagging while minimizing resonator loading.

The spin-wheel (SW) frequency is generated by a set of vertical electric-field
plates distributed around the ring, which provide a common-mode field to the
counter-rotating beams. A feedback loop may be able to stabilize the SW frequency using one
beam as a reference, while the second beam is analyzed in parallel. Independent
I/Q and phase-locked-loop (PLL) estimators extract the instantaneous SW
frequencies $\omega_{\mathrm{SW}}^{\mathrm{CW}}$ and
$\omega_{\mathrm{SW}}^{\mathrm{CCW}}$, referenced to a common probe and clock.
The electric dipole moment (EDM) observable is defined as the differential
frequency
\begin{equation}
\omega_{\mathrm{EDM}} =
\omega_{\mathrm{SW}}^{\mathrm{CW}} -
\omega_{\mathrm{SW}}^{\mathrm{CCW}},
\label{eq:omega_edm}
\end{equation}
which suppresses common-mode drifts and isolates the CP-violating contribution.

In analogy with optical clock comparisons and heterodyne interferometric readout in gravitational-wave detectors, the EDM observable can be extracted from the time evolution of the differential phase between the CW and CCW spin-wheel signals. This approach naturally suppresses common-mode phase noise from the reference oscillator and provides an optimal estimator for the EDM-induced frequency difference.
From this perspective the storage-ring EDM experiment can be regarded as a
``spin interferometer,'' in which the CW beam serves as a reference arm
while the CCW beam constitutes the measurement arm.
The EDM manifests itself as a differential phase drift between the two
arms, analogous to the phase shift produced by a gravitational wave in a
laser interferometer.
This analogy highlights the conceptual unity between coherent polarimetry,
precision frequency metrology, and interferometric sensing techniques.

Recent measurements at COSY~\cite{JEDI:2016swi} and detailed lattice
studies~\cite{Omarov_2022_symmetric} have demonstrated spin-coherence times of
order \(10^3\)~s using RF bunching and optimized sextupole settings.
In the present framework, however, the controlled spin-wheel operation
may substantially modify the effective spin-decoherence dynamics.
As originally discussed by Koop~\cite{Koop:2013vja,Koop:2015lez} and in Sec.~\ref{sec:RF_SCT},
the continuous coherent winding/unwinding of the spin motion can, in
principle, suppress the accumulation of dephasing errors and extend the
effective coherent observation time well beyond the presently
demonstrated scale.
The sensitivity estimates presented in this work therefore include
benchmark coherent integration times up to \(10^5\)~s, not as
experimentally established values, but as motivated long-term operating
scenarios whose ultimate realization will require dedicated lattice,
cooling, and spin-dynamics studies.
Spin-echo techniques~\cite{Hahn1950} can provide an additional extension of the
effective coherence window by reversing controlled spin precession at
well-defined intervals, thereby refocusing residual dephasing,
see also~\cite{derbenev1998radio}.

Taken together, these elements define a coherent polarimetry architecture
that is intrinsically compatible with long spin coherence times and
statistically optimal slope-based estimation.

Beyond searches for a static proton EDM, the same framework extends
naturally to oscillating EDMs and axion-induced spin precession.
In general the axion frequency is not known a priori and must be scanned.
However, if an oscillation frequency is identified---for example through
a haloscope measurement of the axion--photon coupling---the search for
an oscillating nucleon EDM becomes a targeted, single-frequency
measurement.

For a QCD axion, the defining interaction is the gluonic coupling
\begin{equation}
\mathcal{L} \supset
\frac{a}{f_a}\frac{\alpha_s}{8\pi}
G^a_{\mu\nu}\tilde G^{a\,\mu\nu},
\end{equation}
which induces an oscillating CP-violating term and therefore an
oscillating nucleon EDM at frequency $\omega_a = m_a$~\cite{Graham:2013_PhysRevD.88.035023, Graham:2021_PhysRevD.103.055010}.
Observation of such an EDM would directly confirm the gluonic coupling
and distinguish a QCD axion from a generic axion-like particle.

Once the frequency is fixed, coherent demodulation and linear
phase-slope estimation become statistically efficient and saturate the
Cram\'er--Rao bound in the high-SNR regime.
In this sense, coherent polarimetry represents a direct transfer of
resonant, phase-sensitive measurement concepts to storage-ring physics,
with ultimate sensitivity limited by achievable spin coherence time
and phase noise rather than by destructive spin readout.

In addition, recent theoretical developments have also suggested
that nonstandard axion dynamics, including
gravitationally generated or non-compact axions, may
naturally lead to residual strong CP violation at levels
potentially accessible to future proton EDM
experiments~\cite{Karananas2025,Karananas2025_noncompact}.

\section{Signal model: vertical polarization and EDM-induced slope}

In storage-ring EDM searches operated near the frozen-spin condition~\cite{Farley_PhysRevLett.93.052001,Omarov_2022_symmetric}, a nonzero EDM
induces a torque proportional to the effective radial electric field experienced by
the beam.
In the absence of additional controlled spin manipulation, this torque produces a
slow, coherent buildup of vertical polarization.
In the small-angle regime relevant to precision measurements, one may write
\begin{equation}
P_y(t) = P_y^{(0)} + \omega_{\mathrm{EDM}}\, t ,
\label{eq:Py_linear}
\end{equation}
where $P_y^{(0)}$ is determined by initial spin preparation and static imperfections,
and $\omega_{\mathrm{EDM}}$ is proportional to the EDM and the effective radial electric field.

In the present architecture, however, a controlled vertical spin precession
(``spin wheel'') is deliberately introduced~\cite{Koop:2013vja,Koop:2015lez}.
The spin wheel generates a well-defined oscillatory vertical polarization
\begin{equation}
P_y(t) =
P_0 \sin\!\left[
(\omega_{\mathrm{sw}}+\omega_{\mathrm{EDM}}) t + \phi_0
\right],
\label{eq:Py_sw}
\end{equation}
where $\omega_{\mathrm{sw}}$ is the controlled spin-wheel frequency and
$\omega_{\mathrm{EDM}}$ is the much smaller EDM-induced frequency shift.
The EDM therefore appears not as a DC vertical offset but as a small
EDM-induced frequency shift producing a secular phase accumulation.

An important systematic consideration is the possible difference between the
average vertical electric fields experienced by the CW and CCW beams in the
spin-wheel sections. To suppress such effects, the CW and CCW beams are
stored simultaneously and traverse the same electrostatic elements under
nearly identical conditions. The vertical electric-field sections generating
the spin wheel are expected to be located near the ends of the bending
sections and arranged such that the counter-rotating beams pass through the
same vertical-field region simultaneously. This configuration substantially
reduces sensitivity to temporal drifts, power-supply fluctuations, and slow
mechanical or thermal variations of the applied vertical field.

Additional systematic-error suppression is obtained by repeating the
measurement for multiple controlled spin-wheel frequencies generated by
different vertical electric-field strengths. Representative operating points
include
\[
f_{\rm SW}=\{-10,-3,-1,-0.3,-0.1,0.1,0.3,1,3,10\}\ {\rm Hz}.
\]
In practice, the optimal spin-wheel operating frequencies will be determined
experimentally by balancing phase sensitivity, orbit stability, spin
coherence, and systematic-error suppression. Representative operating
points are expected to lie in the approximate range
$f_{\rm SW}\sim \pm(0.1\text{--}10)\ {\rm Hz}$.
A genuine EDM signal should remain independent of the chosen spin-wheel
frequency, whereas many false signals associated with residual field
imperfections or coupling effects are expected to vary with the applied
spin-wheel configuration. Requiring the extracted EDM signal to remain flat
under systematic reversal and scaling of the spin-wheel frequency therefore
provides a powerful internal consistency check.

For representative operation at a spin-wheel frequency of
$f_{\rm SW}\sim 1~{\rm Hz}$, the required ring-averaged vertical
electric field is modest,
\[
E_y^{\rm avg}\sim 9~{\rm V/m},
\]
corresponding to an equivalent magnetic field scale of order
$\sim 20~{\rm nT}$ for $\beta\simeq0.6$. The spin-wheel system is
assumed to consist of approximately 96 localized vertical-field
sections, each about 40~cm long and positioned near the ends of the
bending regions. The corresponding local electric fields are therefore
only of order ${\cal O}(10^2)~{\rm V/m}$.

An important systematic consideration is the spatial uniformity of the
applied vertical electric field. Because the magnetic quadrupoles occupy
only a small fraction of the ring circumference (there are 48 of them with $\pm 0.21 $~T/m, 40~cm long each)~\cite{Omarov_2022_symmetric}, the equivalent orbit
perturbation associated with the spin-wheel field is estimated to remain
at the few-micron level, which is comfortably below the nominal
$\sim10~\mu{\rm m}$ CW--CCW orbit-overlap specification of the storage
ring. The spin-wheel system is therefore not expected to significantly
disturb the baseline orbit-control requirements.

Using this conservative orbit scale together with a target relative
CW--CCW field matching at the $10^{-9}$ level for
$10^{-29}~e\cdot{\rm cm}$ sensitivity implies a tolerable residual field
nonuniformity of order
\[
\frac{\partial E_y}{\partial y}\sim 10^{-3}~{\rm V/m^2},
\]
which is experimentally very modest for centimeter-scale electrostatic
structures. In addition, operation with both positive and negative
spin-wheel frequencies, combined with CW/CCW differencing, provides
further strong suppression of residual static nonuniformities and
associated false-EDM contributions. 

\subsection*{Representative beam and readout parameters}
For concreteness, we summarize the baseline machine, beam, and readout scales
used throughout this work in Tables~\ref{tab:scales} and \ref{tab:params_sym}.
The stored beam consists of $N_b\simeq 80$ bunches per direction with
$N_p\simeq 1.2\times10^8$ protons per bunch and polarization $P\simeq 0.8$.
The coherent polarimeter is operated at a selected bunch-train harmonic
$f_0\simeq 18.18~\mathrm{MHz}$ with loaded quality factor $Q_L\sim 10^5$,
corresponding to a bandwidth $\Delta f \sim f_0/Q_L$ and an effective response time
$\tau \sim Q_L/(\pi f_0)$.
These scales define the effective baseband sampling interval for statistically
efficient phase-slope estimation, while the controlled spin-wheel tone lies in the
range $f_{\mathrm{sw}}\sim \pm(0.1$--$10)~\mathrm{Hz}$ and is fully transmitted in the
demodulated phase envelope. We emphasize that the loaded quality factor
$Q_L\sim10^5$ discussed here refers to the narrowband
coherent-polarimeter (CP) readout resonator used for
phase-sensitive detection of the spin-dependent pickup
signal, and not to the primary storage-ring RF cavity
responsible for longitudinal beam bunching. The CP resonator operates as a weakly
coupled narrowband detection element at the selected
beam harmonic and therefore experiences substantially
different beam-loading and operational constraints than
the main RF accelerating system.

\begin{table*}[t]
\caption{\label{tab:scales}
Characteristic time and frequency scales in the coherent polarimeter concept.}
\centering
\begin{tabular}{l c l}
\hline\hline
Quantity & Symbol & Representative value / comment \\
\hline
LC-resonator carrier frequency (bunch harmonic) & $f_0$ & $18.18~\mathrm{MHz}$ \\
Loaded resonator quality factor & $Q_L$ & $10^{5}$ \\
Resonator bandwidth & $\Delta f$ & $\Delta f \simeq f_0/Q_L \sim 180~\mathrm{Hz}$ \\
Resonator response time (memory) & $\tau$ & $\tau \simeq Q_L/(\pi f_0)\sim 1.8~\mathrm{ms}$ \\
Spin-wheel reference frequency & $f_{\mathrm{sw}}$ & $\pm (0.1-10)~\mathrm{Hz}$ \\
Time between CW and CCW bunch crossings & $T_b$ & $27.5~\mathrm{ns}$ \\
Helicity alternation period (same direction) & $T_h$ & $\sim 2.2~\mu\mathrm{s}$ \\
Spin coherence time & $\mathrm{SCT}\, (T)$ & $10^{3}$--$10^{5}~\mathrm{s}$ \\
Independent sample spacing & $\tau_s$ & $\tau_s \gtrsim \tau$ (baseline $\sim$ ms) \\
\hline\hline
\end{tabular}
\end{table*}

\begin{table*}[t]
\caption{\label{tab:params_sym}
Baseline beam parameters and symmetry-tagged channel construction.}
\centering
\begin{tabular}{l c l}
\hline\hline
Item & Symbol & Meaning / role \\
\hline
Protons per bunch & $N_p$ & $\sim 1.2\times 10^{8}$ \\
Number of stored bunches per direction & $N_b$ & $\sim 80$ \\
Stored polarization & $P$ & $\sim 0.8$ \\
Beam speed & $\beta$ & $\sim 0.6$ \\
Beam--electrode distance & $r$ & $\sim 2~\mathrm{cm}$ \\
Electrode gap & $d$ & $\sim 4~\mathrm{cm}$ \\
Ring circumference & $L$ & $\sim 800~\mathrm{m}$ \\
\hline
Left--right differencing & $V_\Delta = V_L - V_R$ & suppresses common-mode charge pickup \\
Helicity subtraction & $\phi_{+}-\phi_{-}$ & rejects helicity-even pickup/drifts \\
CW--CCW subtraction & $\phi^{\mathrm{CW}}-\phi^{\mathrm{CCW}}$ & rejects direction-even systematics \\
Spin-wheel synchronous detection & $\cos(\Omega_{\mathrm{sw}}t)$ & projects onto the controlled reference tone \\
Slope estimator scaling & $\sigma_b$ & $\sigma_b \propto T^{-3/2}$ (under CR conditions) \\
\hline\hline
\end{tabular}
\end{table*}

Because the experiment is performed in a symmetric hybrid storage ring
configuration~\cite{Omarov_2022_symmetric}, the spin wheel is implemented by
applying a controlled common vertical electric field to both the clockwise (CW)
and counter-clockwise (CCW) beams.
The resulting vertical Lorentz force is balanced by the magnetic focusing system,
so that the closed orbits remain stable while the spins experience a controlled
vertical precession.

Since the two beams circulate in opposite directions, the radial magnetic
fields associated with the bending and focusing lattice are opposite for CW
and CCW motion.
Consequently, the magnetic contribution to the spin-wheel precession frequency
changes sign between the two beams, leading to vertical spin rotations in
opposite directions.

By contrast, the EDM-induced torque arises from the radial electric field of
the electrostatic deflectors.
This radial electric field is identical for both circulation directions, and
the EDM-induced precession therefore enters with the same sign for CW and
CCW beams.
The EDM effect is thus common-mode at the spin-dynamics level, whereas the
electromagnetic signal induced in the pickup electrodes and resonator changes
sign between CW and CCW beams because the beam velocity reverses direction.
As a result, the EDM-induced signal appears with opposite sign in the measured
CW and CCW frequency channels, while several charge-induced backgrounds cancel
under subtraction.

This symmetry separation enables the extraction of the EDM signal from the
difference of the measured spin-wheel frequencies while providing additional
suppression of charge-induced backgrounds and associated systematic effects.

In the coherent polarimeter, the circulating bunched
beam produces a narrowband carrier at the selected
revolution harmonic ($f_0\simeq18.18$~MHz), which is
enhanced by a high-$Q$ LC resonator continuously driven
by a coherent probing tone at the same frequency. The
spin-dependent pickup signal produces a small phase
modulation of this carrier,
\[
V(t)=V_0\cos[\omega_0 t+\delta\phi(t)],
\]
where $\delta\phi(t)$ contains the slow collective spin
dynamics. During spin-wheel operation, the dominant
component of $\delta\phi(t)$ oscillates at the controlled
spin-wheel frequency $f_{\rm SW}\sim\pm(0.1$--$10)$~Hz,
producing narrow sidebands at
$\omega_0\pm\omega_{\rm SW}$ around the carrier.
The resonator discussed here is the weakly coupled
narrowband readout resonator used exclusively for
coherent detection of the spin-dependent pickup signal.

The resonator output is read out using standard IQ
(in-phase and quadrature) demodulation with a local
oscillator phase-locked to the probing tone at $f_0$.
Mixing with the reference oscillator and subsequent
low-pass filtering remove the rapid carrier oscillation
at $\omega_0$, leaving a slowly varying complex
baseband envelope. The resulting baseband spectrum
therefore consists primarily of the narrow spin-wheel
tone and its associated phase fluctuations in the
sub-Hz to $\sim10$~Hz region, together with residual
technical phase noise and any remaining low-frequency
backgrounds.

The experimentally relevant EDM observable is then
obtained from the extracted phase slope, equivalently
the small frequency difference between CW and CCW
spin-wheel oscillations.

The directly measured quantity is the phase time series
$\phi(t)$ of the demodulated resonator signal.
The EDM observable is extracted from the corresponding
phase slope, equivalently from the small frequency
difference between the CW and CCW spin-wheel
oscillations.

Any constant offset in $P_y(t)$, overall signal amplitude variation, or static gain mismatch contributes only to the intercept of the phase fit and does not affect the extracted $\omega_{\mathrm{EDM}}$.
The EDM measurement is intrinsically based on a phase-slope difference between CW and CCW beams, rendering it robust against static offsets and slow drifts in the readout chain.

The hierarchy of symmetry operations described in
Table~\ref{tab:symmetry} is expected to provide
substantial suppression of charge-dominated pickup
backgrounds, potentially reaching the
$10^{13}$--$10^{17}$ level relative to the raw charge
field amplitude.
After this suppression, the residual charge contribution enters primarily as
additive noise near the selected harmonic.
The EDM observable is isolated in symmetry space by:
(i) helicity-odd response,
(ii) frequency selection at $\omega_{\mathrm{sw}}$,
and (iii) the common-mode behavior of the EDM term under CW/CCW reversal.

Because the measurement reduces to estimating a small frequency difference
between two narrowband phase oscillations, statistical sensitivity is governed
by phase-noise properties rather than by absolute signal amplitude.
Under conditions where the per-sample signal-to-noise ratio is sufficient
to ensure unbiased phase estimation and the noise correlations are short-ranged,
the Cram\'er--Rao bound applies, yielding the familiar $T^{-3/2}$ scaling for
frequency (slope) estimation derived in Methods.

\section{Order-of-magnitude signal scale and observable}
\label{sec:Order_of_magn}

Before discussing implementation details, it is useful to establish the physical
scale of the spin-dependent signal and of the observable that is ultimately measured.

A vertically polarized relativistic proton bunch carries a collective magnetic
moment.  In the laboratory frame, the associated near magnetic field produces a
transverse electric field through the motional relation
$\mathbf{E}_{\rm spin}\sim \mathbf{v}\times\mathbf{B}_{\rm spin}$.
Approximating the polarized bunch as a magnetized line
source distributed over an effective longitudinal bunch
length $L_{\rm eff}\sim {\cal O}(1~{\rm m})$,
 the magnetic moment per unit length is
\begin{equation}
m' \simeq \frac{N_p P \mu_p}{L_{\rm eff}},
\end{equation}
and the corresponding near field at transverse distance $r$ scales as a line dipole,
\begin{equation}
B_{\rm spin}(r) \sim \frac{\mu_0}{2\pi}\,\frac{m'}{r^2},
\qquad
E_{\rm spin}(r) \sim \beta c\, B_{\rm spin}(r).
\label{eq:Espin_scaling}
\end{equation}
Here $L_{\rm eff}$ characterizes the effective
longitudinal bunch extent contributing coherently to the
near-field signal and should not be confused with the bunch spacing, or pickup-electrode
length.
For representative storage-ring parameters ($N_p \sim 10^8$, $P \sim 0.8$,
$r \sim 2\,\mathrm{cm}$, and $L_{\rm eff}\sim 1\,\mathrm{m}$), see Table~\ref{tab:params_sym}, this yields a characteristic scale
\begin{equation}
E_{\mathrm{spin}} \sim 10^{-13}\,\mathrm{V/m}.
\end{equation}

By contrast, the charge-induced field from the same bunch is many orders of magnitude
larger.  Modeling the bunch as a relativistic line charge with linear density
$\lambda \simeq N_p e/L_{\rm eff}$, the transverse electric field scale is
\begin{equation}
E_{\rm ch}(r)\sim \frac{\lambda}{2\pi\varepsilon_0 r}
\simeq \frac{N_p e}{2\pi\varepsilon_0\, r\, L_{\rm eff}},
\label{eq:Echarge_scaling}
\end{equation}
which for the same representative parameters is
\begin{equation}
E_{\rm ch}\sim 10\,\mathrm{V/m}.
\end{equation}
Thus, at the raw-field level, the charge-induced contribution exceeds the
spin-dependent contribution by a factor of $\sim 10^{14}$, depending on
geometry and $L_{\rm eff}$.

For the representative scaling estimates used here we
adopt $L_{\rm eff}\sim1$~m following the treatment of
Ref.~\cite{Omarov_2022_symmetric}. Since both the
spin-dependent and charge-induced fields scale in the
same way with $L_{\rm eff}$, the precise numerical choice
does not materially affect the dynamic-range ratio
$E_{\rm spin}/E_{\rm ch}$ emphasized in the present work.

\section{Charge discrimination, noise model, and EDM reach}
\label{sec:noise_reach}

The dominant experimental challenge of non-intercepting polarimetry is dynamic range:
at the pickup electrodes the charge-induced electromagnetic response exceeds the
spin-dependent response by many orders of magnitude
(Sec.~\ref{sec:Order_of_magn}).
The EDM observable is therefore constructed in \emph{symmetry space} rather than at
the raw-field level.
Our analysis channel is defined to be simultaneously:
(i) left--right odd (geometric differencing),
(ii) helicity odd (bunch-level tagging),
(iii) narrowband around the controlled spin-wheel tone, and
(iv) common-mode under CW/CCW reversal for the EDM contribution.
Terms even under these projections do not contribute coherently to the EDM estimator;
after these reversals, any residual charge leakage is treated primarily as additive
narrowband noise rather than as a coherent bias.

\subsection*{Resonant transduction and phase observable}

The pickup electrodes are coupled in differential mode
to a narrowband high-$Q$ superconducting LC resonator
tuned to a selected bunch-train harmonic
($f_0\simeq18.18~\mathrm{MHz}$, $Q_L\sim10^5$).
The resonator acts as a weakly coupled readout element
for the beam-driven carrier at this harmonic. A stable
external reference tone at the same frequency is used
for phase locking and IQ demodulation of the resonator
output. The spin-dependent pickup signal appears as a
small phase modulation of the resonantly enhanced
beam harmonic.
The measured observable is the complex phasor near $f_0$; the spin dynamics appear as a
slow phase modulation of this driven carrier rather than as an energy-buildup process.

After IQ demodulation to baseband, the phase record for each circulation direction
may be written as
\begin{equation}
\phi_j(t)=\phi_{j,0}+\omega_j t+\delta\phi_j(t),\qquad j\in\{\mathrm{CW,CCW}\},
\label{eq:phi_model_linear}
\end{equation}
where $\omega_j$ is the extracted phase slope (angular frequency) of the spin-wheel
oscillation in that beam direction, and $\delta\phi_j(t)$ is stochastic phase noise.
The EDM estimator is formed from the CW/CCW difference,
\begin{equation}
{\omega}_{\rm EDM}
=\frac{1}{2}\left({\omega}_{\rm CW}-{\omega}_{\rm CCW}\right),
\label{eq:omegaEDM_est}
\end{equation}
so that the opposite-sign magnetic spin-wheel contributions cancel while the EDM
contribution adds.

\paragraph{Resonator correlation time and statistically independent sampling.}
The resonator sets the correlation time (memory) of the demodulated phase fluctuations.
For a single-pole response the ringdown time is
\begin{equation}
\tau_c \simeq \frac{Q_L}{\pi f_0}.
\label{eq:tauc_def}
\end{equation}
For $Q_L=10^5$ and $f_0=18.18~\mathrm{MHz}$ this gives
\begin{equation}
\tau_c \simeq \frac{10^5}{\pi(18.18\times10^6)}\simeq 1.75~\mathrm{ms}.
\label{eq:tauc_num}
\end{equation}
Phase samples taken at intervals shorter than $\tau_c$ are correlated by the resonator
memory. We therefore define an effective independent sampling interval $\tau_s$ by
\begin{equation}
\tau_s \gtrsim \tau_c,
\qquad
\text{(baseline: }\tau_s\simeq\tau_c\simeq 1.75~\mathrm{ms}\text{)}.
\label{eq:taus_choice}
\end{equation}
In practice, $\tau_s$ should be taken as the measured correlation time of the final
demodulated phase stream (including any additional digital filtering).
The resonator frequency, probing tone, and coupling
network are assumed to be phase locked and stabilized
well below the effective resonator bandwidth
$f_0/Q_L$, consistent with standard narrowband RF
frequency-control techniques~\cite{PLL_ADI}.

This architecture shares important features with
phase-sensitive resonant detection schemes developed in
narrowband axion experiments, in which a weak coherent
perturbation produces a small phase modulation of a
narrowband resonator and the signal is extracted from
the demodulated baseband phase
record~\cite{ProbingPRD2023,CARAMEL_PRD2026}.
In the present storage-ring application, the dominant
technical backgrounds are expected to arise from thermal
Johnson noise, longitudinal beam phase fluctuations,
beam shot noise, residual charge-induced pickup leakage,
microphonics, and phase instability of the resonant
readout chain. 

For representative parameters
($T_{\rm PE}\simeq300$~K,
$T_{\rm res}\simeq4$~K,
$R=50~\Omega$,
and coupling coefficient $\eta\sim0.1$),
the thermal voltage-noise spectral density coupled from
the room-temperature pickup electrodes is reduced by
the weak coupling according to
\[
\sqrt{S_V^{\rm PE}}
\sim
\eta\sqrt{4k_B T_{\rm PE}R}
\sim
10^{-10}~{\rm V}/\sqrt{\rm Hz}.
\]
This contribution is comparable to the intrinsic thermal
noise of the cryogenic resonator itself operating at
$T_{\rm res}\sim4$~K. Consequently, the total effective
thermal-noise level remains of order
$\sim10^{-10}~{\rm V}/\sqrt{\rm Hz}$, corresponding to
an integrated rms noise of order $\sim10^{-9}$~V within
the resonator bandwidth
$\Delta f\sim f_0/Q_L\sim180$~Hz discussed below.

The principal experimental challenge is the enormous
dynamic range between the coherent charge-induced beam
signal and the spin-dependent component. Prior to any
symmetry operations, the effective field ratio is
estimated to be of order
$E_{\rm ch}/E_{\rm spin}\sim10^{14}$, far beyond the
linear operating range of any practical narrowband RF
readout chain. The coherent polarimeter therefore relies
on substantial front-end suppression before the signal
enters the resonator and amplifier chain. The combined
left--right subtraction, helicity tagging, spin-wheel
frequency selection, and CW/CCW differencing are
expected to reduce the coherent charge background by
many orders of magnitude, potentially reaching the
$10^{13}$--$10^{17}$ level summarized in
Table~\ref{tab:symmetry}. Consequently, the effective
dynamic-range requirement presented to the resonant
detector may be reduced from the raw $\sim10^{14}$ level
to approximately the $10^2$--$10^6$ range, where
operation with modern cryogenic narrowband RF and
phase-sensitive detection systems appears plausible.

Using the representative parameters discussed above,
the spin-dependent resonant signal amplitude is expected
to lie approximately in the nanovolt to sub-microvolt
range depending on the effective coupling, resonator
gain, and coherent integration time, while the raw
beam-driven carrier at the selected harmonic may reach
volt-scale or larger amplitudes before symmetry
suppression. The coherent polarimeter therefore relies
critically on performing substantial common-mode
rejection {\it prior} to the high-gain resonant amplification
stage. After the sequence of symmetry projections, the
remaining narrowband residual background near the
spin-wheel sidebands is expected to approach the thermal
and technical phase-noise floor of the cryogenic readout
system.

\subsection*{Spin-projection noise and standard quantum limit (SQL)}

The fundamental statistical limit of a polarization measurement arises
from spin-projection noise of the stored ensemble. Consider a beam of
$N_{\rm stored}$ spin-$1/2$ particles with collective spin operator

\begin{equation}
\mathbf{S}=\sum_{i=1}^{N_{\rm stored}} \mathbf{s}_i .
\end{equation}

For a beam polarized purely along the longitudinal
direction, the coherent transverse collective spin signal
vanishes,
$\langle S_y\rangle=0$.
Nevertheless, quantum spin fluctuations remain finite,
with
\begin{equation}
(\Delta S_y)^2 = \frac{N_{\rm stored}}{4}.
\end{equation}
The coherent polarimeter is sensitive only to the
collective transverse polarization component generated
by the EDM-induced and spin-wheel-driven spin motion.

Defining the normalized transverse polarization observable

\begin{equation}
P_y = \frac{2 S_y}{N_{\rm stored}},
\end{equation}
the standard quantum limit (SQL) for a single measurement of the
transverse polarization component becomes

\begin{equation}
\sigma_{P_y}^{\rm(SQL)} =
\frac{1}{\sqrt{N_{\rm stored}}}.
\label{eq:SQL}
\end{equation}

For the baseline bunch pattern ($N_b\simeq80$ bunches per direction,
two directions stored) and $N_p=1.2\times10^{8}$ protons per bunch,
\[
N_{\rm stored}\simeq 2N_bN_p \simeq 1.92\times10^{10},
\]
which gives
\[
\sigma_{P_y}^{\rm(SQL)} \simeq 7\times10^{-6}.
\]
In the resonant probing scheme the demodulated phase
of the resonator signal is proportional to the vertical
polarization component of the beam. For small signals,
the corresponding phase fluctuation may therefore be
written as
\[
\sigma_\phi \simeq P\,\sigma_{P_y},
\]
or equivalently
\[
\sigma_{P_y} \simeq \frac{\sigma_\phi}{P},
\]
where $P$ is the beam polarization magnitude.

Using the SQL polarization uncertainty derived above,
the corresponding SQL-equivalent phase uncertainty per
statistically independent sample becomes
\begin{equation}
\sigma_\phi^{(\mathrm{SQL})}
\simeq
\frac{P}{\sqrt{N_{\rm stored}}}.
\end{equation}
For the representative parameters considered here
($P\simeq0.8$ and $N_{\rm stored}\simeq1.92\times10^{10}$),
this gives
\[
\sigma_\phi^{(\mathrm{SQL})}
\sim 6\times10^{-6}\ {\rm rad}.
\]
The SQL fluctuation discussed above should be
understood as the intrinsic collective spin fluctuation
of the stored polarized beam rather than directly as the
measured resonator voltage noise itself. The corresponding
observable phase or voltage fluctuation at the resonator
output depends on the transfer function of the pickup and
readout system. In practice, the equivalent SQL-induced
voltage fluctuation may be estimated by applying the
signal-chain scaling relations derived below
[e.g., Eq.~(\ref{eq:signal_chain})] with the polarization
amplitude replaced by the SQL polarization fluctuation
$\sigma_{P_y}^{\rm(SQL)}$.

The phase-noise estimates quoted above correspond to
approximately one statistically independent resonator
sample, i.e., to an effective averaging time of order
$1/\Delta f\sim 2$~ms determined by the resonator
bandwidth. Coherent averaging over longer integration
times improves the effective phase sensitivity
approximately as $1/\sqrt{T\Delta f}$.

An EDM produces a slow coherent spin rotation away from the horizontal
plane. In the small-angle limit the vertical polarization evolves as
\begin{equation}
P_y(t) \simeq P\,\omega_d\, t ,
\end{equation}
where $P$ is the beam polarization magnitude and $\omega_d$ is the
EDM-induced spin-precession rate.
If one were to estimate the EDM signal from a single measurement of
$P_y$ accumulated over an interval $T$, equating the signal amplitude
to the SQL uncertainty would give
\begin{equation}
\omega_d^{\rm(min)} \sim
\frac{\sigma_{P_y}^{\rm(SQL)}}{P\,T}
\sim
\frac{1}{P\sqrt{N_{\rm stored}}\,T}.
\label{eq:edm_min}
\end{equation}
Equation (\ref{eq:edm_min}) is written in terms of the polarization
observable $P_y$. In the coherent resonant readout,
however, the experimentally measured quantity is the
demodulated phase $\phi(t)$, which is proportional to
the vertical polarization according to
$\sigma_\phi \simeq P\,\sigma_{P_y}$.

In the coherent polarimeter, the EDM observable is obtained
from the slope of a continuous phase time series rather than from a
single endpoint measurement. For statistically independent samples
separated by $\tau_s$, optimal least-squares estimation of the slope
yields the familiar $T^{-3/2}$ scaling of coherent frequency metrology,
\begin{equation}
\sigma_{\omega} \sim
\sqrt{12}\,
\frac{\sigma_{P_y}^{\rm(SQL)}\sqrt{\tau_s}}
{P\,T^{3/2}}
=
\frac{\sqrt{12}\sqrt{\tau_s}}
{P\sqrt{N_{\rm stored}}\,T^{3/2}}.
\end{equation}
This expression highlights the two independent sources of statistical
gain: the $\sqrt{N_{\rm stored}}$ enhancement arising from the number
of polarized particles in the stored beam, and the $T^{3/2}$ improvement
that results from coherent phase tracking over long observation times.

\subsection*{Instrumental phase noise}

Instrumental noise enters the measurement as additive complex voltage
noise on the demodulated carrier. For small fluctuations, voltage noise
maps approximately to phase noise according to
\begin{equation}
\sqrt{S_\phi^{\rm inst}}
\sim
\frac{\sqrt{S_V^{\rm inst}}}{V_c},
\label{eq:Sphi_from_SV}
\end{equation}
where $V_c$ is the rms carrier amplitude at the readout point.
Additional phase-noise contributions may arise from
microphonic motion of the high-$Q$ resonator and mechanical
vibrations of the PE electrode plates, frequency jitter of the reference oscillator,
and residual longitudinal beam phase fluctuations.

In the probing readout architecture~\cite{ProbingPRD2023,CARAMEL_PRD2026}
the detector is designed so that instrumental phase noise remains well
below the spin-projection limit,
\begin{equation}
\sigma_{\phi}^{\rm inst} \ll \sigma_{\phi}^{\rm(SQL)},
\end{equation}
ensuring that the dominant uncertainty per independent sample is set
by quantum fluctuations of the stored spin ensemble rather than by
electronics noise.

For reference, the thermal voltage noise of a resistive port at
temperature $T$ is given by the Johnson--Nyquist spectral density
\begin{equation}
\sqrt{S_V^{\rm th}} = \sqrt{4k_B T R}.
\end{equation}
At $T=4~{\rm K}$ and $R=50~\Omega$ this corresponds to
\[
\sqrt{S_V^{\rm th}} \simeq 1\times10^{-10}\,
{\rm V}/\sqrt{\rm Hz}.
\]
The rms noise voltage depends on the effective measurement bandwidth
$\Delta f$,
\begin{equation}
V_{\rm noise}=\sqrt{4k_BTR\,\Delta f}.
\end{equation}
In the present experiment the bandwidth is set by the resonator,
\[
\Delta f=\frac{f_0}{Q},
\]
which for $f_0=18.18\,{\rm MHz}$ and $Q=10^5$ gives
\[
\Delta f\simeq180\,{\rm Hz}.
\]
The resulting Johnson noise voltage is therefore
\[
V_{\rm noise}\simeq1.3\times10^{-9}\,{\rm V}.
\]
For a modest probe amplitude $V_c\sim10\,{\rm mV}$ this corresponds to
\[
\sigma_\phi^{\rm inst}\sim10^{-7}\,{\rm rad}
\]
per statistically independent sample. This is more than two orders of
magnitude below the SQL-equivalent phase noise derived above.

The SQL-equivalent phase fluctuations discussed here
represent the physical spin fluctuations after they are
converted into the measured resonator phase signal by
the pickup and readout system. By contrast, the
instrumental phase noise originates from additive
detector and readout fluctuations superimposed on the
carrier measurement process. Increasing the probing
carrier amplitude $V_c$ therefore improves the precision
with which the carrier phase is measured and suppresses
the equivalent instrumental phase noise referred to the
measured phase observable. However, the underlying
spin-projection fluctuations correspond to genuine
physical fluctuations of the collective spin signal and do
not disappear simply by increasing the carrier power.

In the present application the probing architecture is
used primarily to provide a stable coherent carrier and a
high-precision phase-sensitive readout of the spin-induced
modulation. The full variance-estimator formalism
discussed in Ref.~\cite{ProbingPRD2023} is not explicitly
required for the present sensitivity estimates.

In this regime the detector noise lies below the quantum fluctuations of
the collective spin ensemble, and the experiment therefore operates in
the spin-projection-noise--limited regime.

A key feature of the probing architecture is that the carrier amplitude
$V_c$ is externally controllable through the injected probing power.
Increasing the probing power improves the phase sensitivity according to
Eq.~(\ref{eq:Sphi_from_SV}) by reducing the equivalent instrumental phase
noise referred to the measured phase observable.

By contrast, the spin-projection noise originates from quantum fluctuations
of the collective spin ensemble and is independent of the injected carrier
amplitude. Consequently, the probing method allows the instrumental phase
noise floor to be pushed below the SQL-equivalent phase fluctuations.

This principle was discussed in the context of axion detection in
Ref.~\cite{ProbingPRD2023}, where sufficiently large probing power allows
the detector to approach the quantum-limited regime. In the present
application, because the SQL-equivalent phase fluctuations are expected to
remain above the quantum noise floor near the $\sim20~\mathrm{MHz}$ operating
frequency, the probing architecture provides a realistic path toward
spin-projection-noise--limited operation.

\paragraph{Dynamic range and symmetry rejection.}

At the raw-field level the electromagnetic field generated by the beam
charge exceeds the spin-dependent field by approximately
\[
\frac{E_{\rm charge}}{E_{\rm spin}} \sim 10^{14}.
\]
The SQL uncertainty of the transverse polarization measurement is
\begin{equation}
\sigma_{P_y}^{\rm(SQL)} \sim \frac{1}{\sqrt{N_{\rm stored}}}
\simeq 7\times10^{-6},
\label{eq:SQL_inst}
\end{equation}
which corresponds to resolving a spin signal at the level of roughly
$\sim 10^{-19}$
of the raw charge-induced field.

However, the experiment does not attempt to resolve the raw
$10^{14}$ charge-to-spin field ratio directly.
Instead, the measurement architecture removes the dominant charge
contribution through a sequence of symmetry projections.

The first stage occurs before the resonator.
The pickup signals undergo left–right subtraction, followed by helicity-tagged subtraction, in which signals
from opposite beam helicities are combined with opposite
signs. Since the dominant charge-induced pickup is
approximately independent of helicity whereas the
spin-dependent contribution reverses sign under
helicity reversal, this operation strongly suppresses
residual charge backgrounds while preserving the
spin-sensitive signal.
These operations must be performed prior to the resonator because the
high-$Q$ resonator responds slowly: its rise time
\[
\tau_r \sim \frac{Q}{\pi f_0} \approx 1.7\,{\rm ms}
\]
is many orders of magnitude longer than the bunch spacing. The resonator
therefore integrates signals over many bunch passages.

The resonator then provides narrowband filtering.
The controlled spin-wheel (SW) modulation shifts the spin signal to a
well-defined frequency in the 0.1--10~Hz range, separating it from
residual DC pickup and other slowly varying backgrounds.
The spin-dependent field therefore appears as a phase modulation of the
stable probing carrier.

Finally, the EDM observable is obtained from the difference between the
spin-wheel frequencies of the clockwise and counter-clockwise beams,
$\omega_{\rm CW}-\omega_{\rm CCW}$, which removes remaining
direction-even contributions.

In this way the resonator never has to accommodate the full
$10^{14}$ charge-to-spin dynamic range: the dominant charge response is
rejected by symmetry before resonant detection, and any residual leakage
enters primarily as additive noise rather than as a coherent background.

\subsection*{Cram\'er--Rao regime and slope estimation}

The EDM observable is extracted from the slope of the demodulated
phase time series. Because the resonator acts as a narrowband filter,
successive phase samples become statistically independent only on
timescales comparable to the resonator rise time
$\tau_s\simeq\tau_c\simeq Q/(\pi f_0)$.
The phase record therefore consists of approximately
$N\simeq T/\tau_s$ statistically independent samples over a coherent
observation interval $T$.

Least–squares (LS) estimation of the slope approaches the
Cramér–Rao (CR) bound provided the usual regularity conditions hold:
(i) the additive phase noise is stationary and approximately Gaussian,
(ii) noise correlations decay on timescales shorter than $\tau_s$,
(iii) phase wrapping and cycle slips are negligible, and
(iv) the signal model consists of a single narrowband tone with slowly
varying phase after the symmetry projections and filtering around the
spin-wheel frequency.

Under these conditions the slope uncertainty follows the familiar
$T^{-3/2}$ scaling characteristic of coherent frequency metrology.

\paragraph{Slope uncertainty.}

For $N$ phase samples $\phi_k=\phi(t_k)$ at uniform spacing
$t_k=k\tau_s$ over a coherent interval $T\simeq N\tau_s$,
and white per-sample phase noise $\sigma_\phi^2$,
the LS/CR variance of the slope estimator is
\begin{equation}
\mathrm{Var}(\omega)
=
\frac{12\sigma_\phi^2}{N(N^2-1)\tau_s^2}
\simeq
12\sigma_\phi^2\frac{\tau_s}{T^3},
\qquad (N\gg1),
\end{equation}
so that
\begin{equation}
\sigma_\omega
\simeq
\sqrt{12}\,
\frac{\sigma_\phi\sqrt{\tau_s}}{T^{3/2}}.
\label{eq:phase_time}
\end{equation}
This $T^{-3/2}$ dependence reflects the fact that the EDM signal is
encoded in a frequency (phase-slope) estimator rather than in a single
endpoint measurement of polarization.

Equation (\ref{eq:phase_time}) shows that for fixed per-sample phase
noise $\sigma_\phi$, increasing the resonator memory
time $\tau_s$ reduces the number of statistically
independent samples available within a fixed coherent
measurement interval $T$. Since $\tau_s\sim Q_L/f_0$,
an arbitrarily large resonator quality factor would
therefore eventually degrade the slope-estimation
sensitivity.

In practice, however, the effective phase noise itself
depends on the resonator bandwidth and improves as the
resonator suppresses broadband detector noise and
enhances the coherent narrowband response. Increasing
$Q_L$ therefore produces two competing effects:
improved instantaneous phase sensitivity but increased
sample correlation time. The optimal operating point is
determined by balancing these effects together with
practical considerations such as resonator stability,
microphonics, beam loading, and technical phase noise.

\paragraph{Conversion to EDM sensitivity.}

In the frozen–spin small-angle limit the EDM-induced spin-precession
rate is
\begin{equation}
\omega_{\rm EDM}=\frac{2dE_{\rm eff}}{\hbar},
\end{equation}
which gives~\cite{Omarov_2022_symmetric,KimSemertzidis2021PRD}
\begin{equation}
\sigma_d=\frac{\hbar}{2E_{\rm eff}}\sigma_\omega .
\end{equation}
Combining this relation with the CR-limited slope uncertainty
gives the approximate SQL-limited scaling
\begin{equation}
\sigma_d^{\rm(SQL)}
\sim
\frac{\hbar}{2E_{\rm eff}}
\frac{\sqrt{\tau_s}}
{\sqrt{N_{\rm stored}}\,T^{3/2}} .
\end{equation}

\paragraph{Baseline SQL-limited estimate.}

Using the nominal parameters
\[
f_0 = 18.18~{\rm MHz},\qquad
Q_L = 10^5,
\]
the resonator memory time is
\[
\tau_s\simeq\tau_c\simeq1.75~{\rm ms}.
\]

For $N_{\rm stored}\simeq1.92\times10^{10}$ 
the spin-projection SQL corresponds to a polarization uncertainty
\[
\sigma_{P}^{\rm(SQL)}\simeq 7 \times10^{-6}.
\]
Since the phase observable in the probed readout is expressed in
polarization-equivalent units, the corresponding SQL level for the
demodulated phase per statistically independent sample is
\[
\sigma_\phi\simeq 7\times10^{-6}.
\]
Taking $E_{\rm eff}=3.4~{\rm MV/m}$ and the SQL phase noise
$\sigma_\phi\simeq7\times10^{-6}$ with sampling interval
$\tau_s\simeq1.75~{\rm ms}$, a coherent interval
$T=10^5~{\rm s}$ yields
\[
\sigma_\omega \approx 4\times10^{-14}\ {\rm rad/s},
\]
which corresponds to a per-fill EDM sensitivity of
\[
\sigma_d^{\rm(per\,fill)} \approx 4\times10^{-34}\ e\cdot{\rm cm}.
\]

\paragraph{100-day running sensitivity.}

For a total live time
\[
T_{\rm run}=100~{\rm days}=8.64\times10^6~{\rm s},
\]
the number of statistically independent fills is approximately
\[
N_{\rm fills}\simeq\frac{T_{\rm run}}{T},
\]
so that the combined statistical sensitivity improves as

\[
\sigma_d^{(100{\rm d})}
=
\frac{\sigma_d^{\rm(per\,fill)}}{\sqrt{N_{\rm fills}}}.
\]

Table~\ref{tab:SCT} summarizes the projected EDM sensitivity for several
assumed spin-coherence times (SCT), which determine the usable coherent
interval $T$. The per-fill sensitivity follows the
$T^{-3/2}$ scaling of the Cramér–Rao bound, while the combined
sensitivity improves as $1/\sqrt{N_{\rm fills}}$ with the number of fills.

\begin{table}[h]
\centering
\caption{Projected EDM sensitivity assuming SQL-limited phase noise
for 100 days of running. The per-fill sensitivity follows the
$T^{-3/2}$ scaling of the Cramér--Rao bound, while the total
experiment sensitivity improves further as $1/\sqrt{N_{\rm fills}}$.}
\label{tab:SCT}
\begin{tabular}{crcc}
\hline
SCT ($T$) & $N_{\rm fills}$ & $\sigma_d^{\rm(per\,fill)}$ & $\sigma_d^{\rm(total)}$ (100 days) \\
\hline
$10^3$ s & $8,600$ & $\sim 4\times10^{-31}$ e$\cdot$cm & $\sim 4\times10^{-33}$ e$\cdot$cm \\
$10^4$ s & $860$ & $\sim 1\times10^{-32}$ e$\cdot$cm & $\sim 3\times10^{-34}$ e$\cdot$cm \\
$10^5$ s & $86$ & $\sim 4\times10^{-34}$ e$\cdot$cm & $\sim 4\times10^{-35}$ e$\cdot$cm \\
\hline
\end{tabular}
\end{table}

\paragraph{Conditions for reaching the quoted sensitivity.}

The sensitivity estimates above rely on several experimental conditions.

First, the symmetry projections described earlier (left--right pickup
subtraction, helicity tagging, harmonic selection, and CW/CCW beam
comparison) must suppress the dominant charge-induced pickup so that
the remaining signal entering the resonator is dominated by the
spin-dependent component. Any residual charge pickup should be
uncorrelated with the spin-wheel phase and therefore contribute only as
additive noise to the demodulated phase record.

Second, the correlation time of the demodulated phase noise should not
exceed the effective sampling interval,
$\tau_s\simeq\tau_c$, ensuring that successive samples are
statistically independent.

Third, the phase tracking must remain stable over the coherent
integration interval, avoiding cycle slips or phase wrapping.

Finally, the instrumental phase noise must remain below the
spin-projection (SQL) level,
$\sigma_\phi^{\rm(SQL)}\simeq7\times10^{-6}$ per statistically
independent sample.

In the probing readout architecture~\cite{ProbingPRD2023,CARAMEL_PRD2026,CARAMEL2025Preprint}
this requirement is readily satisfied.
For typical cryogenic RF noise levels at a $50\,\Omega$ port
($\sqrt{S_V}\approx10^{-10}\,\mathrm{V}/\sqrt{\mathrm{Hz}}$ at
$4\,\mathrm{K}$), achieving SQL-limited phase precision requires a probe
carrier amplitude
\[
V_c \gtrsim 2\times10^{-4}\,\mathrm{V},
\]
corresponding to a probing power of order
\[
P_{\rm probe}\gtrsim10^{-9}\,\mathrm{W}.
\]
This minimum power requirement is extremely small. The estimate $V_c\gtrsim2\times10^{-4}$~V corresponds
to the approximate minimum probe carrier amplitude
required for SQL-limited operation under the assumed
thermal-noise conditions. The representative value
$V_c\sim10$~mV used earlier is simply a convenient
illustrative operating point lying comfortably above
this minimum requirement.
In practice the
probing power can be increased substantially—if needed to optimize the
electro–optic (EO) phase readout—without affecting the SQL-limited
measurement principle. Consequently the instrumental phase noise can be
kept comfortably below the spin-projection noise of the stored beam,
ensuring that the experiment operates in the spin-noise–limited regime
rather than being limited by detector imprecision. Increasing the probe amplitude further reduces instrumental phase noise,
but does not improve the ultimate sensitivity once the spin-projection
limit is reached. Higher probe power may nevertheless be advantageous
for optimizing EO phase detection chain.

\section{Outlook and broader impacts}

The coherent, non-destructive polarimetry framework developed here
establishes a qualitatively new operating regime for storage-ring
precision experiments.
By treating the stored beam polarization as a continuous,
phase-coherent dynamical variable rather than as an ensemble of rare
scattering events, the method eliminates the efficiency and
analyzing-power penalties that historically constrained polarimetry.
Instead of sacrificing stored intensity to obtain intermittent
polarization samples, the experiment continuously tracks a collective
spin observable and extracts the EDM signal from a phase or
frequency slope estimator.
As a result, the full statistical weight of the stored beam can be
exploited over long durations, with sensitivity determined primarily by
phase noise and achievable spin coherence time.

In the context of storage-ring EDM searches, this approach is naturally
matched to operating modes that employ controlled spin precession
(spin wheel), spin-echo sequences, and active beam cooling. Together these techniques motivate coherent integration times in the
range $10^{4}$--$10^{5}\,\mathrm{s}$, where optimal slope estimators provide
substantial statistical gain~\cite{Schmitt2017T32}.
The coherent polarimeter is designed to operate linearly and stably
over such macroscopic time scales, enabling these gains to be realized
in practice.

Relative to conventional scattering polarimetry, where only a small
fraction $\kappa\sim1\%$ of the stored beam participates in the
measurement and the analyzing power is $A\simeq0.6$, non-intercepting
readout provides an immediate utilization gain
\begin{equation}
G_{\rm coh/scatt}=
\frac{1}{\sqrt{\kappa\,A^2}}
\approx 17 .
\end{equation}
More importantly, coherent polarimetry produces a continuous phase
record over the full coherent interval $T$, allowing the EDM observable
to be extracted from within-fill slope fitting rather than from a
single accumulated asymmetry.
The statistical advantage therefore grows parametrically with the
coherent observation time $T$.

For the baseline parameters considered in this work, projection-noise
limited operation with spin-coherence times approaching
$T\sim10^{5}\,\mathrm{s}$ yields a frequency sensitivity of order
\[
\sigma_\omega \sim {\rm few}\times10^{-14}\ \mathrm{rad/s}.
\]
Using the relation
\[
\omega_{\rm EDM}=\frac{2dE_{\rm eff}}{\hbar},
\]
and an effective radial electric field
$E_{\rm eff}\simeq3.4\,\mathrm{MV/m}$,
this corresponds to a per-fill EDM sensitivity at the level of
\[
\sigma_d^{\rm(fill)}
\sim {\rm few}\times10^{-34}\,e\!\cdot\!\mathrm{cm}
\qquad (T\sim10^{5}\,\mathrm{s}).
\]

Averaging over $\mathcal{O}(100)$ days of running,
corresponding to many statistically independent fills of duration $T$,
improves the statistical reach by $1/\sqrt{N_{\rm fill}}$,
bringing the projected sensitivity into the
$\sim10^{-34}\,e\!\cdot\!\mathrm{cm}$ range under SQL-limited
assumptions.

The spin-wheel concept and the coherent phase
polarimeter are complementary but logically distinct
elements of the present proposal. The spin wheel itself
already provides important advantages for storage-ring
EDM experiments, including controlled modulation of the
EDM observable away from DC, improved rejection of
slow systematic drifts, and frequency-based signal
extraction. These benefits remain relevant even when
using a conventional destructive polarimeter.

The coherent phase polarimeter, on the other hand,
provides a non-destructive resonant readout capable of
continuous phase tracking with high efficiency and
potentially much improved statistical sensitivity through
long coherent integration times. The full suppression and
sensitivity gains discussed in this work arise from the
combination of both concepts operating together.

\paragraph{Potential systematic errors.}

The present work should be viewed primarily as a
conceptual and analytical framework establishing the
basic symmetry structure, coherent phase-readout
principles, and projected scaling behavior of the
spin-wheel coherent polarimeter. Although the proposed
suppression chain is strongly motivated by the underlying
CW/CCW, helicity, and phase-sensitive symmetries, a
complete validation of the achievable suppression factors
will ultimately require detailed spin-tracking,
beam-dynamics, RF-system, and electromagnetic
simulation studies together with dedicated prototype
measurements.

Particularly important future investigations include
systematic studies of geometric-phase effects, residual
CW/CCW asymmetries, orbit and centroid stability,
vertical-field matching, synchrotron and RF-induced
modulations, helicity imbalance, non-ideal resonator features,
and long-term phase stability. A comprehensive
systematic-error and tolerance budget for the full
spin-wheel/coherent-polarimeter configuration will be an
essential component of future experimental design
studies.

\paragraph{Broader impact and paradigm shift.}

More broadly, coherent non-intercepting polarimetry changes the
operating mode of storage-ring spin experiments from occasional
polarization estimates per fill to continuous phase-coherent tracking,
analogous to frequency metrology.
This enables substantially longer coherent integration, makes
slope-based extraction statistically optimal, and removes the intensity
and efficiency penalties inherent to scattering polarimetry.

The same architecture is directly applicable to a wide range of
precision programs, including polarized-proton and light-ion rings at
the EIC, muon $g\!-\!2$–style rings, and storage-ring searches for EDMs
and oscillating spin-precession signals in deuterons, ${}^3$He, and
antiprotons.
By extending storage-ring EDM sensitivity toward (and potentially into)
the Standard-Model range—a regime not yet reached for any particle—this
approach motivates renewed quantitative engagement from lattice-QCD
calculations needed to interpret such measurements. In this sense, coherent polarimetry enables storage-ring spin experiments to operate in a regime analogous to modern frequency metrology, where the ultimate sensitivity is determined by quantum fluctuations of the measured ensemble rather than by detector efficiency.

\paragraph{Experimental outlook.}

A particularly important proof-of-principle milestone
would be simultaneous operation of the coherent
polarimeter together with a conventional destructive
polarimeter. Such operation would permit direct
cross-calibration of the extracted spin dynamics,
spin-wheel frequency, phase evolution, and systematic
reversals between the two methods under identical beam
conditions. The coherent polarimeter could then be
validated progressively while maintaining an established
independent polarization measurement system during
commissioning and early experimental operation.

\section{Methods}
\label{sec:methods}

\subsection{Differential phase readout and the spin interferometer analogy}
\label{subsec:spin_interferometer}

The coherent polarimetry scheme developed in this work can be viewed as
the spin-dynamics analogue of a heterodyne interferometer.
In interferometric metrology, two coherent oscillators are compared
through their phase evolution, and the observable of interest is
encoded in the time dependence of the phase difference.
This principle underlies many precision measurements, including optical
frequency metrology and gravitational-wave detectors, where the
differential phase of two coherent fields is tracked with high
precision using digital phasemeters.

In the present experiment the clockwise (CW) and counter-clockwise
(CCW) spin-wheel signals play the role of the two coherent oscillators.
Their phases can be written as
\begin{equation}
\phi_{CW}(t)=\omega_{CW}t+\phi_{CW,0}, \,\,
\phi_{CCW}(t)=\omega_{CCW}t+\phi_{CCW,0}.
\end{equation}
The experimentally relevant observable is the differential phase
\begin{equation}
\Delta\phi(t)=\phi_{CCW}(t)-\phi_{CW}(t).
\end{equation}
Substituting the expressions above gives
\begin{equation}
\Delta\phi(t)=(\omega_{CCW}-\omega_{CW})t+\Delta\phi_0,
\end{equation}
where $\Delta\phi_0$ is the initial phase difference.
The time derivative of the differential phase therefore yields the
frequency difference between the two spin-wheel signals,
\begin{equation}
\frac{d\Delta\phi}{dt}=\omega_{CCW}-\omega_{CW}.
\end{equation}

In the spin-wheel configuration the EDM signal appears as a differential
frequency shift between the CW and CCW beams,
\begin{equation}
\omega_{CCW}-\omega_{CW}=2\,\omega_{EDM}.
\end{equation}
The EDM observable is therefore obtained from the slope of the
differential phase,
\begin{equation}
\omega_{EDM}=\frac{1}{2}\frac{d}{dt}\Delta\phi(t),
\end{equation}
which defines the frequency estimator used throughout this work.

This measurement strategy is conceptually identical to heterodyne
interferometry in gravitational-wave detectors such as LIGO and LISA,
where the science signal is obtained from the phase evolution of a
beat note between two coherent optical fields.
Modern gravitational-wave phasemeters track this phase continuously
using digital phase-locked loops, achieving sub-$\mu$rad phase
precision in the millihertz-to-hertz frequency band
\cite{Gerberding2015,Gerberding2013}.
In that context the interferometer output is reconstructed entirely
from the measured phase time series.

A key advantage of using the differential phase is the suppression of
common-mode phase noise.
Fluctuations of the reference oscillator or electronics that enter both
CW and CCW channels appear as common-mode phase variations and cancel
to first order in $\Delta\phi(t)$.
This principle closely parallels the common-mode noise rejection
strategies used in gravitational-wave interferometry, including the
time-delay interferometry combinations developed for space-based
detectors to suppress laser frequency noise
\cite{Tinto2014}.

The statistical properties of such frequency-difference measurements
are well understood from the theory of frequency standards.
The stability of an oscillator comparison is commonly characterized
using the Allan variance formalism introduced in the context of atomic
clocks and precision frequency metrology \cite{Allan1966}.
In that framework the relevant observable is the evolution of the
phase difference between two oscillators as a function of averaging
time.
For a coherent sinusoidal signal the Cramér--Rao bound implies that
the uncertainty of the frequency estimator scales as
\begin{equation}
\sigma(\omega)\propto T^{-3/2},
\end{equation}
where $T$ is the total coherent observation time.
The EDM observable therefore inherits the same $T^{-3/2}$ scaling,
characteristic of phase-coherent frequency measurements.

In this sense the coherent polarimeter operates in a regime analogous
to modern frequency metrology, where the ultimate statistical
sensitivity is determined by the quantum fluctuations of the measured
ensemble rather than by detector efficiency.

\subsection{Frequency resolution of the phase--locked CW spin--wheel signal}
\label{sec:CW_resolution}

In the proposed scheme the clockwise (CW) spin--wheel signal is actively
phase locked to the same reference oscillator that is used as the probing
signal for the CW and CCW resonators.
The feedback loop adjusts the spin--wheel drive (for example through the
Wien-filter amplitude or phase) so that the CW phase follows the reference
oscillator.
Let the CW spin signal be
\begin{equation}
s_{CW}(t)=A\cos(\omega_{CW}t+\phi_{CW}),
\end{equation}
while the reference oscillator is
\begin{equation}
r(t)=A_r\cos(\omega_{\rm ref}t).
\end{equation}
The phase detector measures the phase difference
\begin{equation}
\Delta\phi(t)=\phi_{CW}(t)-\omega_{\rm ref}t ,
\end{equation}
and the feedback loop drives this quantity toward zero.
After lock acquisition the CW phase can therefore be written as
\begin{equation}
\phi_{CW}(t)=\omega_{\rm ref}t+\phi_n(t),
\end{equation}
where $\phi_n(t)$ represents the residual phase noise of the locked system.

The CW frequency is estimated from the evolution of this phase record
over a coherent interval $T$.
For a coherent sinusoidal signal with rms phase uncertainty
$\sigma_\phi$, the Cram\'er--Rao bound for frequency estimation gives

\begin{equation}
\sigma(\omega_{CW})
\simeq
\frac{\sqrt{12}\,\sigma_\phi}{T^{3/2}},
\end{equation}
or equivalently
\begin{equation}
\sigma(f_{CW})
\simeq
\frac{\sqrt{12}\,\sigma_\phi}{2\pi\,T^{3/2}} .
\end{equation}
The achievable frequency resolution therefore depends
primarily on the phase-measurement precision and the
coherent integration time. In practice, this also
requires that the demodulated phase $\phi(t)$ be tracked
with sufficient signal-to-noise ratio on timescales short
compared with the spin-wheel period
$1/f_{\rm SW}$ so that the coherent oscillatory phase
evolution can be reconstructed continuously throughout
the measurement interval. For the representative resonator bandwidth and carrier
levels considered here, the coherent carrier phase can in
principle be tracked on timescales substantially shorter
than the spin-wheel period. The practical short-timescale
signal-to-noise ratio is therefore expected to be limited
primarily by technical phase-noise sources and residual
charge-background leakage rather than by thermal noise
alone.

The feedback loop does not fundamentally change the estimator variance;
rather, it suppresses slow drifts of the CW spin--wheel frequency so that
the phase evolution remains coherent over long time intervals and the
$T^{-3/2}$ scaling derived in the previous subsection can be realized
experimentally.

It is important to emphasize that locking the CW spin--wheel frequency to
the reference oscillator does not suppress the EDM signal.
The EDM torque produces a differential frequency shift between the CW and
CCW beams, and the EDM observable is therefore encoded in the time
evolution of their phase difference.
In symmetry language the feedback acts only on the
\emph{common-mode} spin--wheel frequency,
\begin{equation}
\omega_{\rm CM}=\frac{\omega_{CW}+\omega_{CCW}}{2},
\end{equation}
which is stabilized by the feedback loop.
Operationally, the measurement is closely analogous to
lock-in detection: the spin-wheel motion shifts the EDM
observable into a narrowband modulation channel at
$f_{\rm SW}$, and the resonator phase is demodulated
synchronously with respect to this reference frequency.
The experimentally relevant quantity is the small
difference between the extracted CW and CCW
spin-wheel frequencies, which may be determined through
continuous coherent phase tracking and offline frequency
analysis.

The EDM signal resides in the antisymmetric channel,
\begin{equation}
\omega_{\rm AS}=\frac{\omega_{CCW}-\omega_{CW}}{2}=\omega_{EDM},
\end{equation}
which is unaffected by the locking procedure.

If the CW and CCW channels share the same reference oscillator, phase
noise of the reference largely cancels when forming the EDM observable,
\begin{equation}
f_{EDM}=\frac{1}{2}\left(f_{CCW}-f_{CW}\right),
\end{equation}
so that the differential measurement retains the same statistical
scaling while being largely insensitive to common-mode oscillator noise.

This differential phase-tracking strategy is analogous to techniques used
in precision frequency metrology and interferometric phase measurements,
where coherent phase records are analyzed to obtain high-resolution
frequency estimates.

\subsection{RF pickup and resonant transduction at \texorpdfstring{$f_0=18.18$}{f0=18.18} MHz}
\label{subsec:methods_rf_pickup}

\subsubsection{Resonator design point and bandwidth}
\label{subsubsec:methods_resonator_design}

The pickup electrodes (PE) are located at room temperature and act purely as
broadband field sensors for the beam-induced electromagnetic signal.
Each electrode plate is capacitively coupled to the readout chain, and the
signals from the left and right plates are combined in a hybrid subtractor
that forms their differential output.
This left--right subtraction suppresses the dominant charge-induced component
of the pickup signal while preserving the spin-dependent contribution.

Because the pickup electrodes and front-end coupling
network operate at room temperature, the initial thermal
noise contribution is determined primarily by the
corresponding warm source impedance. Weak capacitive
coupling between the pickup electrodes and the
high-$Q$ resonator reduces the large coherent beam
carrier entering the resonant readout chain and thereby
substantially relaxes the required amplifier dynamic
range and beam-loading constraints. Although this weak
coupling reduces both the signal and the associated
thermal-noise voltage coupled into the resonator, the
measurement remains fundamentally phase sensitive: the
relevant observable is the phase modulation of the large
coherent carrier rather than the direct detection of a
small standalone spin-induced voltage amplitude.
Consequently, high carrier phase sensitivity may still be
maintained even when the coupled signal power is kept
deliberately small.

Although the pickup electrodes themselves operate at
room temperature and therefore introduce a
room-temperature front-end thermal-noise contribution,
the weak pickup--resonator coupling
($\eta\sim0.1$) substantially attenuates this injected
noise before it enters the resonant readout chain.
Because the coupled thermal-noise power scales as
$\eta^2 T$, the effective room-temperature contribution
seen by the resonator becomes comparable to that of a
few-kelvin source. The cryogenic superconducting
resonator therefore remains important for achieving high
loaded quality factor, narrowband filtering, low
internal dissipation, and stable coherent phase readout.
The resonator reduces the effective measurement
bandwidth and improves the achievable phase
sensitivity even though the initial pickup signal
originates from a warm electrode system.

The differential output of the hybrid is then transmitted to the cryogenic
readout stage.
Before reaching the resonant circuit, the signal passes through a fast
electronic switching stage that routes the clockwise (CW) and
counter-clockwise (CCW) signals to their respective resonators.
At the same stage the signal is multiplied by $\pm1$ according to the
helicity tagging of the beam bunches, which provides an additional level
of charge suppression while leaving the spin-dependent signal unchanged.

After this switching and tagging stage the signal is coupled to a
high-$Q$ LC resonator located at approximately $4\,\mathrm{K}$, as
illustrated in Fig.~\ref{fig:Res_PLL}.
The LC circuit therefore provides the first narrowband amplification
stage for the spin-dependent signal, while the pickup electrodes
themselves remain broadband sensors.

Importantly, the capacitance that defines the LC resonance is provided
by the resonator circuit rather than by the pickup electrodes.
The PE plates are therefore electrically isolated from the resonant
circuit except through the controlled capacitive coupling network.
Although the present discussion focuses primarily on a cryogenic
superconducting LC implementation, the coherent polarimeter concept
itself is not restricted to a particular resonator technology. Other
high-$Q$ narrowband resonators, including quartz-crystal-based systems,
may also provide attractive alternatives with potentially excellent
frequency stability, reduced microphonic sensitivity, and substantially
simplified implementation.

In this configuration the resonators are driven by the differential
voltage generated by the pickup electrodes while being simultaneously
excited by a coherent probe signal at frequency
$f_{\mathrm{ref}}=f_0$.
The spin-dependent electromagnetic field of the polarized beam
phase-modulates the resonator response, producing sidebands around the
carrier that encode the spin-wheel dynamics.

\subsection{Effect of RF momentum oscillations on the spin--wheel coherence}
\label{sec:RF_SCT}

Operation of the spin wheel (SW) requires the application of a small
vertical electric field $E_y$ around the ring. This field produces a
vertical force $qE_y$ which is compensated by the Lorentz force generated
by a radial magnetic field $B_r$ from the magnetic focusing system,
\begin{equation}
qE_y = q\,v B_r .
\end{equation}
The compensating radial magnetic field therefore depends on the particle
velocity,
\begin{equation}
B_r = \frac{E_y}{v}.
\end{equation}
Synchrotron oscillations driven by the RF system introduce a momentum
modulation
\begin{equation}
\frac{\delta p}{p} \lesssim 5\times10^{-4},
\end{equation}
which induces a velocity variation~\cite{Bovet:1970}
\begin{equation}
\frac{\delta v}{v} \approx
\frac{1}{\gamma^2}\frac{\delta p}{p}.
\end{equation}
Here the term ``RF system'' refers to the conventional
storage-ring longitudinal RF cavities used for beam
bunching and synchrotron confinement, not to the
helicity-tagging electronics or to the resonator probing
tone used in the coherent polarimeter readout.
Fluctuations of the storage-ring RF amplitude and phase
can couple to synchrotron motion and therefore contribute
to low-level phase noise in the demodulated signal.
However, these effects are expected to enter primarily as
common-mode longitudinal beam fluctuations and are
strongly suppressed by the symmetry projections,
CW/CCW differencing, and narrowband synchronous
detection at the controlled spin-wheel frequency.
Residual RF-induced phase fluctuations are treated as
technical noise contributions to the phase-noise budget.

Longitudinal momentum spread and synchrotron motion can
in principle modify the bunch form factor and therefore
produce slow amplitude and phase modulation of the
coherent carrier signal at the selected bunch harmonic.
In practice, however, the stored beam remains RF-bunched
throughout operation, and the relevant longitudinal beam
parameters are stabilized by the conventional storage-ring
RF system and synchrotron-feedback controls.

Because the coherent polarimeter measures the phase of a
narrowband resonantly enhanced carrier rather than the
instantaneous bunch shape itself, moderate variations of
the bunch length primarily enter as common-mode
modulation of the carrier amplitude and phase. Residual
effects are expected to appear as technical phase-noise
contributions and can be monitored continuously through
the bunch spectrum, synchrotron sidebands, and
independent longitudinal beam diagnostics.

For the proton EDM ring ($\gamma \approx 1.25$) this gives
\begin{equation}
\frac{\delta v}{v} \approx 3\times10^{-4}.
\end{equation}
Because the spin precession rate generated by the radial magnetic field
scales with $B_r$, the relative modulation of the spin frequency satisfies
\begin{equation}
\frac{\delta \omega_s}{\omega_s}
\simeq
\frac{\delta B_r}{B_r}
\approx
-\frac{\delta v}{v}.
\end{equation}

The RF oscillation periodically modulates the velocity and therefore the
spin frequency. The linear term averages to zero over many synchrotron
oscillations, but the quadratic term produces a residual spread of the
spin frequency,
\begin{equation}
\frac{\Delta f}{f_{SW}}
\sim
\left(\frac{\delta v}{v}\right)^2
\sim 10^{-7}.
\end{equation}
The resulting absolute spin-frequency spread is therefore
\begin{equation}
\Delta f \sim f_{SW}\times10^{-7}.
\end{equation}
For a spin-wheel frequency of $1\,\mathrm{Hz}$ this corresponds to
$\Delta f \sim 10^{-7}\,\mathrm{Hz}$ and a coherence time
$\tau \sim 10^{7}\,\mathrm{s}$.
For $f_{SW}=100\,\mathrm{Hz}$ the spread increases to
$\Delta f \sim 10^{-5}\,\mathrm{Hz}$, corresponding to
$\tau \sim 10^{5}\,\mathrm{s}$.

These estimates indicate that RF-induced momentum oscillations are
unlikely to limit the spin coherence time at low spin-wheel frequencies,
while at higher spin-wheel frequencies the effect may become comparable
to the targeted spin-coherence time of
$10^{4}$--$10^{5}\,\mathrm{s}$.

In addition to RF-induced momentum oscillations, the spin-frequency spread
also receives contributions from residual radial magnetic fields.
As emphasized by Koop~\cite{Koop:2013vja,Koop:2015lez}, the total spin-tune
spread can be written schematically as the quadratic combination of the
momentum spread and magnetic-field imperfections,
\begin{equation}
\frac{\Delta f}{f_{SW}}
\sim
\sqrt{
\left(\frac{\delta p}{p}\right)^2 +
\left(\frac{\delta B_r}{B_r}\right)^2 } .
\end{equation}

Here the first term represents the effect of momentum-dependent
spin precession, while the second term arises from fluctuations of the
radial magnetic field that balances the vertical electric force.
The dominant contribution to the spin-coherence limit is therefore
determined by whichever of these two mechanisms is larger.

\subsection{Systematic error considerations in the spin--wheel mode}
\label{sec:SW_systematics}

A complete systematic-error analysis for the spin-wheel (SW) mode remains
to be carried out. Nevertheless, it is useful to examine qualitatively
how the dominant systematic effects known from the frozen-spin method are
expected to behave when the SW configuration is employed.

In the frozen-spin scheme the principal systematic sources have been
identified by Omarov and collaborators~\cite{Omarov_2022_symmetric}. These include
(i) vertical electric fields,
(ii) vertical velocity components,
(iii) residual electric focusing, and
(iv) combinations of unwanted fields that generate geometric-phase
effects. In the conventional frozen-spin configuration these effects
are particularly important because the frozen-spin condition corresponds
to a near cancellation of the magnetic-dipole spin precession. As a
result the residual spin motion becomes highly sensitive to small
perturbations of the electromagnetic fields.

The SW method modifies this situation in an important way. In this mode
a small vertical electric field $E_y$ is intentionally applied around
the ring to drive a controlled spin rotation in the vertical plane.
The resulting vertical electric force $qE_y$ is compensated by the
Lorentz force from a radial magnetic field generated by the magnetic
focusing system, so that the beam orbit remains stable while the spin
precesses with a finite frequency. The resulting spin-wheel frequency
typically lies in the range $0.1$--$10~\mathrm{Hz}$.
The EDM observable is extracted from the difference between the SW
frequencies of the two counter-rotating beams.

Because the spin motion is no longer tuned to a near cancellation of
the magnetic-dipole precession, the system is less sensitive to
perturbations that would otherwise generate large false signals in the
frozen-spin configuration.

Vertical electric fields constitute the leading systematic in the
frozen-spin approach because they produce a false EDM signal through
the magnetic-dipole interaction. This effect
is naturally and effectively suppressed by the simultaneous storage of clockwise (CW)
and counter-clockwise (CCW) beams. The EDM observable is obtained from
the difference of the two measured SW frequencies, so common-mode
contributions from vertical electric fields cancel to first order.

Vertical velocity components represent another important systematic
source. If the beam develops a small radial spin component $S_r$, a
vertical velocity $v_y$ in the presence of the radial electric field
$E_r$ produces a motional magnetic field
\[
B_z = -\,\frac{v_y E_r}{c^2},
\]
directed longitudinally. This field couples to the magnetic dipole
moment and can generate a false EDM signal by inducing a vertical
spin-precession rate that mimics the effect of a true EDM. The presence
of a residual radial spin component $S_r$ is essential for this
mechanism, since it allows the longitudinal magnetic field to rotate
the spin toward the vertical direction. In the frozen-spin
configuration this mechanism is particularly dangerous because the
spin dynamics occurs near the cancellation point of the magnetic-dipole
precession. As a result even a small longitudinal magnetic field can
produce a linear accumulation of vertical polarization.

In a highly symmetric storage ring the magnitude of the vertical
velocity is strongly suppressed by the lattice design. Hybrid symmetric
ring configurations have been shown to reduce the leading contributions
from such imperfections by orders of magnitude, see Fig.~(6) of
\cite{Omarov_2022_symmetric}. Residual vertical motion can nevertheless
arise from lattice imperfections such as quadrupole offsets and field
errors. For representative quadrupole misalignments at the
$\sim10~\mu$m level, the resulting coherent orbit excitation may reach
the $\sim 10^2~\mu$m scale, although the associated vertical velocity
$\beta_y$ remains strongly suppressed.

The spin-wheel (SW) mode provides an additional level of suppression.
In this configuration the dominant spin motion is intentionally driven
by the applied vertical electric field together with its compensating
radial magnetic field, which generates a controlled vertical spin
precession. The longitudinal magnetic field arising from the
$v_y \times E_r$ term therefore acts only as a weak perturbation to the
much stronger effective radial magnetic field that governs the spin
rotation. Consequently the associated false EDM signal is reduced by
an additional large factor compared with the frozen-spin case.

Residual electric focusing can also shift the spin-precession frequency.
As in the frozen-spin configuration, these effects can be diagnosed and
minimized through spin-based alignment procedures. In addition, the SW
method provides an independent handle by reversing the applied vertical
electric field, $E_V \rightarrow -E_V$, which flips the sense of the
spin-wheel rotation. True EDM signals remain unchanged under this
reversal, while focusing-induced effects change sign or cancel in the
frequency comparison.

Geometric-phase effects arising from the combination of two or more
unwanted fields are also expected to be reduced in the SW configuration.
Such effects are enhanced in the frozen-spin method because the spin
dynamics occurs near the cancellation point of the magnetic-dipole
precession. Driving the spin motion at a finite SW frequency moves the
system away from this condition and suppresses the associated
geometric-phase contributions.

An additional advantage arises when the underlying systematic sources
are already suppressed by lattice symmetry. In the hybrid symmetric
ring configuration proposed by Omarov {\it et al}~\cite{Omarov_2022_symmetric},
the leading field imperfections are designed to be small from the
outset. In the conventional frozen-spin configuration, false EDM
signals can arise linearly from such imperfections through couplings
between residual fields and the particle motion. In particular, the
motional-field mechanism discussed above leads schematically to
\begin{equation}
d_{\rm false} \propto \beta_y E_r S_r .
\end{equation}
In the SW configuration, once the dominant linear couplings are removed
by the ring symmetry, CW/CCW comparison, and operation away from the
near-cancellation condition of the frozen-spin method, the leading
residual contributions arise only at quadratic order in the remaining
imperfections,
\begin{equation}
d_{\rm false}^{\rm SW} \propto
(\beta_y E_r S_r)^2 .
\end{equation}
Because the quantity $\beta_y$ is already strongly suppressed in the
symmetric-ring design~\cite{Omarov_2022_symmetric}, this quadratic
scaling provides an additional reduction of several orders of magnitude
in the effective systematic error. This feature may open the possibility
of reaching experimental sensitivity levels comparable to those
expected from Standard-Model contributions.

In summary, although a detailed systematic study remains necessary,
the SW configuration provides several intrinsic advantages. It reduces
the sensitivity of the spin dynamics to small perturbations of the
electromagnetic fields, converts several leading systematic effects
from linear to quadratic order in the underlying imperfections, and
introduces additional experimental handles such as $B_r$ reversal and
CW/CCW frequency comparison. Together these features suggest that the
dominant systematic uncertainties may be substantially reduced relative
to the conventional frozen-spin approach.

\subsection{Narrowband resonant response at $f_0=18.18$~MHz}
\subsubsection{Gaussian form factor}
\label{subsubsec:methods_gauss_formfactor}
The LC resonator responds to the Fourier component of the periodic bunch-train
voltage waveform in a narrow band around $\omega_0=2\pi f_0$. For an rms bunch
length $\sigma_z$ and Gaussian longitudinal profile, the bunch form factor at
angular frequency $\omega$ is
\begin{equation}
F(\omega)=\exp\!\left[-\frac{(\omega\sigma_t)^2}{2}\right],
\qquad
\sigma_t=\frac{\sigma_z}{\beta c}.
\end{equation}
For $\sigma_z=0.994$~m and $\beta\simeq0.6$, $\sigma_t\simeq5.52$~ns and
$F(\omega_0)\simeq0.82$ at $f_0=18.18$~MHz.

\subsubsection{Transverse ensemble averaging and betatron enhancement}
\label{subsubsec:methods_betatron_enhancement}

The instantaneous spin-induced field at the pickup electrodes depends on
the transverse displacement of individual particles executing betatron
oscillations. While the beam centroid remains approximately centered
between the pickup plates, individual particles oscillate about the
equilibrium orbit, thereby modulating their distance from the electrodes.

For a particle with horizontal displacement $x$, the distance to the
right pickup electrode becomes $r_0-x$. Since the spin-induced near
field of the polarized bunch approximately follows the scaling of a
line dipole,
\begin{equation}
B_{\rm spin}(r)\propto \frac{1}{r^2},
\end{equation}
the corresponding transverse electric field at the pickup electrode
scales approximately as
\begin{equation}
E_{\rm spin}(x)
=
E_{\rm spin}(r_0)
\left(\frac{r_0}{r_0-x}\right)^2 .
\end{equation}

The collective signal observed by the pickup electrodes is therefore
determined by the ensemble average over the transverse beam
distribution. In a realistic storage ring, the transverse particle
coordinates are expected to follow an approximately Gaussian
distribution determined by the lattice optics and equilibrium
emittance. The corresponding average may therefore be written
schematically as
\begin{equation}
\left\langle \frac{1}{(r_0-x)^2} \right\rangle
=
\frac{1}{\sqrt{2\pi}\sigma_x}
\int_{-\infty}^{+\infty}
\frac{e^{-x^2/(2\sigma_x^2)}}{(r_0-x)^2}\,dx,
\end{equation}
where $\sigma_x$ denotes the rms horizontal beam size.

This averaging enhances the relative contribution of particles passing
closer to the pickup electrodes because of the strong near-field
distance dependence. However, the enhancement is expected to be more
moderate than the idealized hard-edge estimates obtained using uniform
transverse particle distributions. In addition, the effective field at
the electrodes is influenced by image-current redistribution and
boundary-field modifications associated with the conducting vacuum
chamber geometry.

Consequently, the enhancement factor entering the signal-chain
estimate in Eq.~(\ref{eq:signal_chain}) should be interpreted only as
an order-of-magnitude representation of near-field geometric
sensitivity. A complete quantitative determination would require
self-consistent electromagnetic simulations including the beam,
vacuum chamber, pickup electrodes, and resonator coupling network.

For an order-of-magnitude estimate, if the previous
maximum betatron amplitude scale $a=18~{\rm mm}$ is
interpreted as approximately $3\sigma_x$, then
$\sigma_x\simeq6~{\rm mm}$ for $r_0=20~{\rm mm}$.
The corresponding Gaussian average gives a 
modest near-field enhancement,
\begin{equation}
{\cal F}_G
=
r_0^2
\left\langle \frac{1}{(r_0-x)^2}\right\rangle
\sim 1.5 ,
\end{equation}
with the precise value depending on the physical aperture
cutoff and the conducting vacuum-chamber boundary
conditions.

\subsubsection{Bunch-train scaling}
\label{subsubsec:methods_train_scaling}
For a uniformly distributed train of $N_b$ bunches, the complex Fourier component
near $\omega_0$ adds coherently,
\begin{equation}
\widetilde V_{\rm train}(\omega_0)\simeq
N_b\,F(\omega_0)\,\widetilde V_{1,\delta}(\omega_0),
\end{equation}
so the effective enhancement factor per direction is $N_bF\simeq66$ for the
baseline parameters.  The same factor applies to the spin-wheel sidebands
$\omega_0\pm\Omega_{\rm sw}$ for $\Omega_{\rm sw}\ll\omega_0$.
The LC resonator does not measure the instantaneous voltage produced
during a single bunch passage, but the narrowband Fourier component of
the periodic bunch train within its bandwidth
$\Delta f \simeq f_0/Q_L$ centered at $f_0$.
When the resonator is tuned to a harmonic of the revolution frequency,
the bunch train provides a coherent drive at $\omega_0=2\pi f_0$.
The resonator therefore integrates the drive coherently over the
characteristic ring-down time
\begin{equation}
\tau_c = \frac{Q_L}{\omega_0}.
\end{equation}

Equivalently, the steady-state resonant voltage is enhanced relative to
the instantaneous pickup voltage by approximately the loaded quality
factor $Q_L$.  In addition, the Fourier component of the periodic bunch
train contributes coherently through the bunch-train factor
$N_bF(\omega_0)$.

Combining the expressions derived above, the resonant voltage generated
by the spin-dependent electromagnetic field of the polarized beam can be
written compactly as
\begin{equation}
V_{\rm res}
\simeq
Q_L\,N_b\,F(\omega_0)\,
\mathcal{F}\,
\frac{\mu_0}{2\pi}
\frac{\beta c\,N_p P \mu_p}{L_{\rm eff}}
\frac{d}{r_0^2},
\label{eq:signal_chain}
\end{equation}
which explicitly shows the complete signal chain from the proton magnetic
moment to the measurable resonant voltage at the output of the LC
system.  Here $\mathcal{F}$ denotes the enhancement factor arising from
the ensemble averaging over the transverse betatron amplitude
distribution, discussed below in~\ref{subsubsec:methods_betatron_enhancement}.

For clarity, the parameters entering Eq.~(\ref{eq:signal_chain}) and the
representative baseline values used in the present estimate are listed
in Table~\ref{tab:signal_parameters}.
Substituting these representative values into
Eq.~(\ref{eq:signal_chain}) yields a resonant voltage scale of order $10^2$\,nV.
\paragraph{Spin-induced voltage scale.}
\begin{table}[t]
\centering
\caption{Representative parameters used for the signal estimate.}
\label{tab:signal_parameters}
\begin{tabular}{lc}
\hline
Parameter & Value \\
\hline
Number of protons per bunch $N_p$ & $1.2\times10^8$ \\
Beam polarization $P$ & $0.8$ \\
Proton magnetic moment $\mu_p$ & $1.41\times10^{-26}$ J/T \\
Relativistic factor $\beta$ & $0.6$ \\
Pickup distance $r_0$ & $2$ cm \\
Plate separation $d$ & $4$ cm \\
Effective bunch length $L_{\rm eff}$ & $1$ m \\
Loaded resonator $Q_L$ & $10^5$ \\
Bunch-train factor $N_bF(\omega_0)$ & $\approx66$ \\
Betatron enhancement $\mathcal{F}$ & $\approx 1.5$ \\
\hline
\end{tabular}
\end{table}

The differential voltage generated by the spin-dependent field at the
pickup electrodes can be estimated by combining the field scale of
Eq.~(9) with the geometric electrode separation and the coherent
enhancement factors associated with the bunch train, transverse
beam motion, and resonant readout.  The resulting signal scale can
be written as
\begin{equation}
V_{\rm spin}
\simeq
E_{\rm spin}\, d\,
N_b\,
{\mathcal F}\,
Q\,
\label{eq:Vspin_estimate}
\end{equation}
where $E_{\rm spin}$ is the spin-induced electric field at the pickup
location, $d$ is the effective electrode separation, $N_b$ is the
number of stored bunches contributing coherently at the selected
bunch-train harmonic, ${\mathcal F}$ is the betatron form-factor enhancement
associated with the transverse beam oscillation envelope, and $Q$ is the
loaded quality factor of the resonant readout circuit.

Using representative parameters
$E_{\rm spin}\sim10^{-13}\,\mathrm{V/m}$,
$d\simeq4\times10^{-2}\,\mathrm{m}$,
$N_b\simeq66$,
${\mathcal F}\simeq 1.5$, and
$Q\simeq10^{5}$
one obtains
\begin{equation}
V_{\rm spin}\sim10^{-7}\,\mathrm{V},
\end{equation}
corresponding to a spin-induced signal of order
$\sim10^2\,\mathrm{nV}$ at the resonator input.

For practical applications it is useful to express the result as a
normalized scaling relation for the resonator output voltage,
\begin{equation}
\begin{aligned}
V_{\rm sig} \approx
3\,{\rm nV}\,
&\left(\frac{Q_L}{10^5}\right)
\left(\frac{N_b}{66}\right)
\left(\frac{N_p}{1.2\times10^8}\right)
\left(\frac{P}{0.8}\right) 
\left(\frac{\mathcal F}{1.5}\right)\\
&\times
\left(\frac{\beta}{0.6}\right)
\left(\frac{d}{4\,{\rm cm}}\right)
\left(\frac{2\,{\rm cm}}{r_0}\right)^2
\left(\frac{\eta}{0.1}\right),
\end{aligned}
\label{eq:normalized_scaling}
\end{equation}
where $\eta$ denotes the effective pickup--resonator
coupling efficiency. A moderately weak coupling of
order $\eta\sim0.1$ is assumed, ensuring that the
pickup extracts only a small fraction of the beam-induced
electromagnetic energy while still maintaining a
high signal-to-noise ratio in the resonator readout.
The coherent polarimeter operates primarily as a
phase-sensitive probe rather than as a power-extraction
device, so strong beam loading is neither required nor
desirable.
The normalization used in
Eq.~(\ref{eq:normalized_scaling}) includes representative
weak-coupling and geometric overlap factors associated
with the pickup-electrode configuration and resonator
mode matching. Consequently, the resulting coupled
signal amplitude entering the readout chain is reduced
relative to the idealized resonator-scale estimate.

In practice, the effective coupling is determined by the
pickup geometry and by the impedance matching between
the pickup electrodes and the resonator input circuit.
The coupling strength may be tuned using standard RF
methods, including capacitive or inductive coupling
adjustments and control of the external resonator
quality factor.

Equation~(\ref{eq:normalized_scaling}) makes explicit how the signal
depends on the key machine and detector parameters.  In particular,
the signal scales linearly with the resonator quality factor,
the bunch population, and the beam polarization, while the dependence on
the pickup geometry is dominated by the inverse-square scaling with the
distance between the beam and the electrodes.

Thus the polarized bunch train produces a coherent narrowband signal at
$f_0=18.18$~MHz at the nanovolt level.  Although extremely small in
absolute magnitude, such signals are accessible using coherent probing
techniques such as the electro-optic readout employed in the CARAMEL
method~\cite{CARAMEL_PRD2026}, which converts the resonator phase modulation into an optical
sideband measurable with shot-noise-limited sensitivity.

\subsection{Charge pickup estimate and symmetry discrimination}
\label{subsec:methods_charge_symmetry}

\subsubsection{Line-charge scale}
\label{subsubsec:methods_line_charge}

Modeling the bunch as a relativistic line charge with linear density
$\lambda\simeq N_p e/L_{\rm eff}$, the transverse electric field at
distance $r$ is
\begin{equation}
E_{\rm ch}(r)\sim\frac{\lambda}{2\pi\varepsilon_0 r}
\simeq \frac{N_p e}{2\pi\varepsilon_0\,r\,L_{\rm eff}}.
\end{equation}
For representative parameters this yields $E_{\rm ch}\sim 17$~V/m,
many orders of magnitude larger than the spin-dependent field.

In principle, increasing the number of pickup-electrode
stations can enhance the coherent signal and reduce the
impact of instrumental phase noise by providing multiple
independent measurements of the same spin-dependent
phase modulation. This is especially useful during
commissioning and systematic studies, where comparison
among several pickup locations provides important
cross-checks of orbit, alignment, impedance, and local
background effects.

Once the measurement reaches the spin-projection-noise
(SQL) limit of the stored ensemble, however, additional
pickup stations no longer improve the fundamental
statistical sensitivity, since they observe the same
collective spin state. They may nevertheless remain
valuable for redundancy, diagnostics, and suppression of
technical noise below the SQL. In the present work we
therefore treat localized pickup electrodes as the baseline
unit, with the number and placement of such stations to
be optimized in future detailed RF and beam-dynamics
studies.

The coherent polarimeter is designed to measure a
narrowband RF phase modulation associated with the
collective spin dynamics of a relativistic bunched beam.
Electric pickup electrodes couple naturally to the
transverse motional electric field generated by the
spin-dependent beam magnetization and integrate
directly into the resonant phase-sensitive readout
architecture employed here.

By contrast, SQUID-based systems are optimized
primarily for low-frequency magnetic-flux measurements
and become less natural for operation at the MHz
bunch-harmonic frequencies considered in the present
scheme. In addition, the spin-dependent signal is many
orders of magnitude smaller than the charge-induced
beam fields and accelerator RF backgrounds, creating
significant dynamic-range and linearity challenges for
magnetic pickup systems. The differential electric-pickup
geometry provides substantial common-mode rejection
already at the pickup stage, thereby strongly relaxing the
requirements on the downstream resonant and optical
readout system.

\subsubsection{Leakage from charge pickup: centroid offset and helicity rejection}
\label{subsubsec:methods_leakage_budget}

The bunch charge produces a large common-mode signal on the pickup
electrodes. For a centered beam this signal is identical on the two
plates and is rejected by left--right subtraction. The dominant residual
arises from a finite beam centroid offset
$\delta x\equiv\langle x\rangle$ relative to the electrode midpoint at
transverse distance $r_0=d/2$.

Betatron oscillations, even with large amplitudes, do not generate a
differential signal provided the transverse distribution remains
symmetric, since the leading leakage term is odd in $x$ and therefore
proportional only to $\delta x$.

Approximating the bunch as a relativistic line charge with
$\lambda=N_p e/L_{\rm eff}$, the quasi-static potential is
\begin{equation}
\Phi(r)=\frac{\lambda}{2\pi\varepsilon_0}\ln r,
\end{equation}
which leads to a differential voltage
\begin{equation}
V_{\Delta,\mathrm{ch}}
=
\frac{\lambda}{2\pi\varepsilon_0}
\ln\!\left(\frac{r_0-\delta x}{r_0+\delta x}\right).
\end{equation}
For $|\delta x|\ll r_0$,
\begin{equation}
V_{\Delta,\mathrm{ch}}
\simeq
-\frac{\lambda}{2\pi\varepsilon_0}
\left(\frac{2\delta x}{r_0}\right).
\end{equation}

Using representative parameters
$N_p=1.2\times10^8$, $L_{\rm eff}=1$~m,
$r_0=2$~cm, and $\delta x\simeq4~\mu$m,
one finds
\begin{equation}
|V_{\Delta,\mathrm{ch}}|\simeq 140~\mu\mathrm{V},
\label{eq:volt_ref}
\end{equation}
or $\sim280~\mu$V when both CW and CCW beams contribute.

Helicity-tag subtraction rejects this helicity-even charge signal by a
factor $R_h\sim10^4$--$10^5$, yielding a residual
\begin{equation}
V_{\mathrm{ch,res}}
\sim
3\text{--}30~\mathrm{nV}.
\label{eq:Vch_res}
\end{equation}
This sets the relevant starting point for the narrowband SNR analysis at
$f_0=18.18$~MHz.

Because the residual charge pickup still exceeds the spin-induced signal,
direct broadband detection would remain charge dominated. The measurement
therefore relies on narrowband resonant detection, in which the LC circuit
integrates the periodic bunch drive while rejecting off-resonant
backgrounds. In the CARAMEL scheme, a coherent probe converts the
spin-induced phase modulation into a detectable optical sideband with
strong rejection of residual charge pickup.

\paragraph{Hierarchy of symmetry-based charge rejection.}

The measurement employs a sequence of symmetry projections that suppress
the charge-dominated background prior to resonant detection:

\emph{Left--right geometric subtraction.}
The dominant charge signal is common-mode on the two pickup plates and
is rejected by antisymmetric subtraction. The residual differential
signal arises from a finite beam centroid offset $\delta x$ relative to
the electrode midpoint and scales linearly as
$V_{\Delta,\mathrm{ch}}\propto \delta x/r_0$.

In the absence of active correction, the achievable suppression is set
by the natural orbit stability. For the representative value
$\delta x\simeq4~\mu$m adopted here, this corresponds to a rejection
factor of $\sim10^3$--$10^4$ relative to the raw charge signal.

The representative orbit-stability scale
$\delta x\sim4~\mu{\rm m}$ used in Eq.~(\ref{eq:volt_ref})
should be interpreted as an effective residual
closed-orbit fluctuation entering the spin-sensitive
pickup asymmetry after orbit correction and CW/CCW
common-mode suppression. It is not intended to
represent solely the momentum-dependent synchrotron
orbit excursion.

In practice, the residual orbit motion may contain
contributions from momentum offsets, magnet field
errors, alignment imperfections, mechanical vibration,
high-voltage fluctuations of the electrostatic bending
system, and other sources of closed-orbit distortion.
The coherent polarimeter sensitivity depends primarily
on the remaining differential CW/CCW orbit mismatch
at the pickup locations rather than on the absolute
closed orbit itself.

For a representative bending radius of order
$\sim100$~m, high-voltage stability at the
$0.1$~ppm level corresponds to orbit variations of
order $\sim10~\mu{\rm m}$. Stability at this level is
consistent with modern ultra-stable high-voltage
systems developed for precision electrostatic
applications such as PTOLEMY ~\cite{PTOLEMY_HV2026}. The present ring configuration
is assumed to contain 48 electrostatic bending
sections, with separate high-voltage supplies for the
positive and negative electrodes, corresponding to
approximately 96 independently regulated channels.
To the extent that residual fluctuations among these
channels are only partially correlated, the resulting
global closed-orbit perturbation is expected to be
further reduced by statistical averaging. In practice,
the relevant requirement involves the combined
stability of the electrostatic system, including voltage
stability, mechanical alignment, and vibration control
of the bending electrodes.

\emph{Helicity-tag subtraction.}
Charge pickup is helicity-even, while the spin-dependent signal reverses
sign under helicity inversion. The achievable rejection is therefore
set by the equality of the effective bunch populations associated with
opposite helicities.

In the operating scenario considered here, the injected beam is first
allowed to debunch and expand, forming a nearly uniform distribution
around the ring. A second-harmonic RF system is then applied to rebunch
the beam symmetrically, producing pairs of bunches with nearly equal
intensity. An RF solenoid (or equivalent spin rotator) is used to rotate
the spin from the vertical into the horizontal plane, with opposite
rotation phases applied to successive bunches. This procedure generates
a train of bunch pairs with equal populations and opposite helicities,
separated by the helicity period $T_h$.

Because the beam originates from a common, uniformly distributed parent
population, the statistical uncertainty in the relative populations of
the two helicity states is set by counting statistics. For a total stored
intensity of order $N_{\rm tot}\sim10^{10}$ particles, the fractional
population imbalance is therefore
\begin{equation}
\frac{\Delta N}{N}\sim\frac{1}{\sqrt{N_{\rm tot}}}\sim10^{-5},
\end{equation}
up to small additional contributions from capture and RF manipulation.

Since the charge-induced pickup is proportional to the bunch population,
this residual imbalance directly limits the cancellation of the
helicity-even signal. The corresponding rejection factor is therefore
of order $10^4$--$10^5$, which we adopt as a conservative estimate.

Conceptually, the helicity-tagging system acts as a
synchronous modulation scheme analogous to lock-in
detection: the spin-dependent component reverses sign
under helicity reversal, whereas the dominant
charge-dependent electromagnetic background remains
approximately unchanged. The resonator and subsequent
phase-sensitive demodulation therefore project the signal
onto the controlled helicity/spin-wheel modulation
channel, strongly suppressing static and slowly varying
common-mode backgrounds.

\emph{Spin-wheel modulation.}
The spin-wheel (SW) configuration shifts the EDM-sensitive signal from
DC to a well-defined modulation frequency
$f_{\rm SW}\sim 0.1$--$10$~Hz, while the dominant charge pickup remains
quasi-static on this timescale. The achievable rejection is therefore
set by the ratio of the readout phase-noise spectral density at low
frequency to that at $f_{\rm SW}$,
$S_\phi(f\!\sim\!0)/S_\phi(f_{\rm SW})$.

In the electro-optic probing scheme considered here, the resonator is
interrogated by a coherent optical carrier at $f_0=18.18$~MHz, and the
phase readout is expected to be shot-noise limited above the
low-frequency technical-noise region. As in standard heterodyne and
balanced homodyne measurements, the phase-noise spectrum exhibits a
$1/f$-like rise at low frequencies due to thermal drifts, electronics
offset fluctuations, and residual laser intensity noise, while
approaching a white (shot-noise-limited) floor above a few Hz.

Operating at $f_{\rm SW}\gtrsim1$~Hz therefore places the signal in the
transition region toward the white-noise plateau, where the noise level
is reduced by typically $10^3$--$10^4$ relative to the quasi-DC regime.
Such suppression factors are routinely achieved in modulated detection
systems that shift the signal away from $1/f$ noise
(e.g., lock-in and optical heterodyne techniques), and are therefore
adopted here as a conservative estimate.

\emph{CW--CCW comparison.}
The EDM observable is constructed from the appropriate combination of the
measured spin-wheel phases or frequencies of the CW and CCW beams. The
key point is not that the radial electric field changes sign between the
two beams; it is common to both. Rather, the imposed spin-wheel rotation
has opposite sign for CW and CCW, whereas many instrumental phase
perturbations---including RF-reference drift, probe-laser phase drift,
slow resonator detuning, and readout gain/offset fluctuations---enter
both channels as nearly common-mode contributions.

The resulting CW--CCW combination therefore acts as a differential
estimator that suppresses correlated technical noise in the same spirit
as common-reference two-channel phase measurements. The attainable
rejection is set by the mismatch between the CW and CCW transfer
functions, not by the absolute size of the technical noise itself. With
a shared RF/optical reference and well-matched pickup/resonator/readout
chains, a fractional mismatch at the $10^{-3}$--$10^{-4}$ level over the
spin-wheel band appears to be a realistic conservative target, implying
an additional suppression factor of order $10^{3}$--$10^{4}$.

\paragraph{Thermal-noise detectability and phase-sensitive readout.}

The pickup electrodes operate at room temperature and
therefore generate thermal Johnson noise associated
with the effective front-end source impedance.
However, the pickup electrodes are coupled only weakly
to the cryogenic high-$Q$ resonator, with a representative
coupling coefficient $\eta\sim0.1$. Consequently, both
the coherent beam-driven carrier and the associated
warm front-end thermal noise are attenuated before
entering the narrowband cryogenic readout chain.

The effective coupled thermal voltage-noise spectral
density may therefore be estimated approximately as
\begin{equation}
v_n^{\rm eff}
\sim
\eta \sqrt{4kTR}.
\end{equation}

For representative values $R=50~\Omega$,
$T=300$~K, and $\eta=0.1$,
\begin{equation}
v_n^{\rm eff}
\simeq
0.09~\mathrm{nV}/\sqrt{\mathrm{Hz}}.
\end{equation}

With $f_0=18.18$~MHz and $Q_L=10^5$, the effective
resonator bandwidth is
\begin{equation}
\Delta f=\frac{f_0}{Q_L}\simeq182~\mathrm{Hz},
\end{equation}
corresponding to an integrated rms thermal noise within
the resonator bandwidth of
\begin{equation}
v_{n,\mathrm{BW}}
=
v_n^{\rm eff}\sqrt{\Delta f}
\simeq1.2~\mathrm{nV}.
\end{equation}

For representative coupled spin-signal amplitudes of
order
\begin{equation}
V_{\rm sig}\sim3~\mathrm{nV},
\end{equation}
the instantaneous narrowband voltage-amplitude
signal-to-noise ratio within the resonator bandwidth is
therefore expected to be of order a few.

The quantity $v_{n,\mathrm{BW}}$ corresponds to the rms
noise within the effective resonator bandwidth and
therefore to a statistically independent sampling time
of order
\begin{equation}
\tau_c \sim \frac{1}{\Delta f}\sim 2~\mathrm{ms},
\end{equation}
comparable to the resonator coherence time.
Coherent averaging over longer observation intervals
improves the effective signal-to-noise ratio approximately
as
\begin{equation}
\mathrm{SNR}\propto\sqrt{T\Delta f},
\end{equation}
where $T$ is the integration time. For $T=1$~s, this
corresponds to an additional improvement factor of
approximately $\sqrt{182}\sim13$.

In practice, however, the coherent polarimeter does
not rely on direct power detection of the small
spin-induced signal. The experimentally relevant
observable is the phase modulation of the much larger
coherent carrier generated at the selected beam
harmonic. The EDM signal is extracted from the
differential phase evolution of the CW and CCW
spin-wheel oscillations through coherent offline
frequency analysis rather than through tight real-time
stabilization of the spin-wheel frequency.

Operationally, the measurement is therefore closely
analogous to lock-in detection: the spin-wheel motion
shifts the EDM observable into a narrowband modulation
channel at $f_{\rm SW}$, and the resonator phase is
demodulated synchronously with respect to this reference
frequency. In order for the spin-wheel oscillation to be
reconstructed reliably, the spin-dependent phase
evolution must remain measurable on timescales shorter
than the inverse spin-wheel frequency $1/f_{\rm SW}$.

The beam coupling associated with the pickup electrodes
is expected to remain small because the coherent
polarimeter operates in a weak-coupling,
phase-sensitive regime rather than as a strongly
beam-loaded RF structure. The pickup electrodes are
geometrically compact, approximately symmetric with
respect to the beam orbit, and coupled only weakly to
the high-$Q$ resonator. Their electromagnetic impact on
the beam is therefore expected to be comparable to or
smaller than that of conventional beam-position-monitor
(BPM) pickups and other passive diagnostic elements
commonly used in storage rings.

Nevertheless, a complete evaluation of the beam
coupling impedance, wakefields, coherent tune shifts,
beam-induced heating, residual charge-background
leakage, longitudinal beam phase noise, microphonics,
and amplifier dynamic-range requirements requires
dedicated electromagnetic simulation and
accelerator-physics analysis. These effects are expected
to scale primarily with the effective pickup impedance
and external coupling strength and can be optimized
through electrode geometry and resonator coupling
design.

\subsection{Statistical uncertainty of slope extraction}
\label{subsec:Statistics}

\paragraph{Scope.}
The EDM observable in the present framework is extracted from the \emph{slope}
(or equivalently, a frequency) of a phase-coherent time series accumulated over a
coherent interval of duration $T$. The key statistical question is therefore:
given a noisy time series, what is the uncertainty with which a linear-in-time
slope can be estimated?

\paragraph{Roadmap of this subsection.}
We begin with the general least-squares / Cram\'er--Rao result for slope estimation
with explicit sampling time $\tau_s$, and highlight the origin of the $T^{-3/2}$
scaling. We then compare coherent (non-destructive) readout to scattering polarimetry
in the two practically relevant regimes:
(i) many independent samples within a fill versus
(ii) one effective estimate per fill.
Finally, we discuss the role of multiple resonators, the SQL and spin squeezing,
and we contrast coherent polarimetry noise with axion-haloscope (thermal-field) noise.

\paragraph{Notation and conventions.}
\begin{table*}[t]
\label{tab:notation}
\centering
\caption{Notation and conventions}
\begin{tabular}{lc}
\hline
Parameter & Explanation \\
\hline
$T$ & total coherent observation duration used in the slope fit \\
$\tau_s$ & effective sampling / averaging interval for one phase estimate \\
$N$ & number of samples in the fit, $N\simeq T/\tau_s$ \\
$x_k$ & measured time series (e.g.\ demodulated phase) at $t_k=k\tau_s$ \\
$b$ & slope parameter (e.g.\ rad/s if $x$ is a phase) \\
$\sigma_x$ & per-sample rms uncertainty of $x_k$ \\
$S_\phi$ & (approximately white) phase-noise PSD of the coherent readout \\
$B$ & effective phase-measurement bandwidth (set by resonator + filtering) \\
$A$ & analyzing power (scattering polarimetry) \\
$\kappa$ & detected fraction of stored beam in scattering mode \\
$N_{\rm stored}$ & number of stored particles contributing coherently \\
$N_{\rm res}$ & number of independent pickup channels (resonators) combined \\
$P_0$ & typical polarization magnitude used to map $P\leftrightarrow \phi$ \\
$f_0$ & resonator probe/carrier frequency (current baseline: $18.18$\,MHz) \\
\hline
\end{tabular}
\end{table*}
Throughout, ``independent samples'' refers to (approximately) uncorrelated \emph{imprecision}
noise in successive bins of duration $\tau_s$; it does not imply fresh particle ensembles. The notation used is shown in Table~\ref{tab:notation}.

\subsubsection{General slope-estimation result (with explicit sampling time)}
\label{subsubsec:Statistics_general_slope}

Consider a time series sampled uniformly at times $t_k = k\tau_s$ over a total
duration $T \simeq N\tau_s$, modeled as
\begin{equation}
x_k \;=\; x_0 + b\,t_k + n_k,
\label{eq:M2_linear_model}
\end{equation}
where $b$ is the slope (units: $1/\mathrm{s}$ if $x$ is a phase), and the noise
samples are independent with
\begin{equation}
\mathrm{E}[n_k]=0,
\qquad
\mathrm{Var}(n_k)=\sigma_x^2.
\label{eq:M2_noise_def}
\end{equation}
Here $\sigma_x$ is the \emph{per-sample rms} of $x$ (e.g.\ radians if $x$ is a phase).

For $N$ equally spaced samples, the least-squares (and Cram\'er--Rao) variance of the
slope estimator is
\begin{equation}
\mathrm{Var}(b)
=
\frac{\sigma_x^2}{\sum_{k=0}^{N-1}(t_k-\bar t)^2}
=
\frac{12\,\sigma_x^2}{N(N^2-1)\,\tau_s^2},
\label{eq:M2_varb_exact}
\end{equation}
with $\bar t$ the sample mean time. For $N\gg 1$ (i.e.\ $T\gg\tau_s$),
\begin{equation}
\mathrm{Var}(b)
\;\simeq\;
\frac{12\,\sigma_x^2\,\tau_s}{T^3},
\qquad
\sigma_b
\;\simeq\;
\sqrt{12}\,\frac{\sigma_x\,\sqrt{\tau_s}}{T^{3/2}}.
\label{eq:M2_sigmab_discrete}
\end{equation}

Equation~\ref{eq:M2_sigmab_discrete} makes explicit that, in the optimal phase-coherent regime,
the statistical reach improves rapidly with integration time as $T^{-3/2}$,
so that for a fixed running time the coherent polarimeter is expected to
achieve a substantially larger integrated SNR than scattering-based readout;
if residual technical noise dominates $\sigma_x$, this per-sample uncertainty
can be further reduced by coherently combining $N_{\rm res}$ independent
pickup channels distributed around the ring, yielding an additional
$\sqrt{N_{\rm res}}$ sensitivity gain until the spin-projection floor is reached.

Here ``spin-projection-noise-limited'' refers to the per-sample uncertainty
$\sigma_x$ of a statistically independent measurement bin of duration $\tau_s$.
Once $\sigma_x$ is reduced to the standard quantum limit of the stored ensemble,
additional pickup channels cannot further lower the noise floor, and the final
slope sensitivity then follows directly from Eq.~\ref{eq:M2_sigmab_discrete} through coherent averaging
over $N\simeq T/\tau_s$ samples.
In practice, the coherent polarimeter does not operate bunch-by-bunch:
each phase sample of duration $\tau_s\gtrsim\tau_c$ corresponds to the
coherently accumulated response of the full stored bunch train (e.g. 80
bunches) over $\sim 10^{4}$--$10^{5}$ passages, so that the relevant SNR is
that of the collective phase estimate per $\tau_s$, not per individual bunch.

\paragraph{Why $T^{-3/2}$ appears.}
The slope uncertainty improves faster than $1/T$ because the estimator uses
\emph{all} samples across the interval:
(i) the linear trend grows as $\propto T$, and
(ii) averaging over $N=T/\tau_s$ independent samples reduces noise as $\propto 1/\sqrt{N}$.
Together this yields $\sigma_b\propto (1/T)\times(1/\sqrt{T})$ in the denominator,
i.e.\ $\sigma_b\propto T^{-3/2}$ (up to the explicit $\sqrt{\tau_s}$ factor).

\paragraph{Continuous-time form (PSD language).}
More importantly, coherent polarimetry provides a continuous, non-intercepting phase
time series over the full coherent interval $T$. In this case the EDM signal is
extracted from a slope fit to a measurement record whose white phase-noise floor is
characterized by a PSD $S_\phi$. The corresponding Cram\'er--Rao bound yields the
standard slope scaling
\begin{equation}
\sigma_{\omega_d}^{\rm coh}
\;\simeq\;
\sqrt{12}\,\frac{\sqrt{S_\phi}}{T^{3/2}},
\label{eq:M2_coh_CRB_PSD}
\end{equation}
independent of the particular discretization interval used in the analysis.

\subsubsection{Comparison with scattering polarimetry}
\label{subsubsec:Statistics_scattering_compare}

\paragraph*{Scattering polarimetry: event-counting statistics.}
A conventional scattering polarimeter infers polarization from an asymmetry in detected
scattering events~\cite{KimSemertzidis2021PRD}.  For a given time gate (or fill) in which $N_{\rm det}$ events are
recorded and the effective analyzing power is $A$, a standard estimator is
\begin{equation}
\hat P \;=\; \frac{1}{A}\,\frac{N_+ - N_-}{N_+ + N_-},
\end{equation}
where $N_\pm$ are event counts in two symmetric detector arms (or spin-sensitive
channels).  For small asymmetries and Poisson counting statistics, the per-estimate
polarization uncertainty is
\begin{equation}
\sigma_{P}^{\rm scat}
\;\simeq\;
\frac{1}{A\sqrt{N_{\rm det}}}.
\label{eq:scat_sigmaP_basic}
\end{equation}
Only a fraction $\kappa\ll 1$ of the stored beam contributes to detected events over the
gate, so $N_{\rm det}\simeq \kappa\,N_{\rm stored}$ (up to acceptance, target thickness,
and running mode).  Equation~(\ref{eq:scat_sigmaP_basic}) therefore embodies the familiar
``utilization $\times$ analyzing-power'' penalty:
\begin{equation}
\sigma_{P}^{\rm scat}
\;\simeq\;
\frac{1}{A\sqrt{\kappa\,N_{\rm stored}}}.
\label{eq:scat_sigmaP_kappa}
\end{equation}

\paragraph*{From polarization estimates to an EDM slope.}
If scattering polarimetry could provide a sequence of statistically independent
polarization (or spin-angle) estimates over the coherent interval $T$---for example,
by operating with many short, weakly intercepting gates---then the same slope-estimation
theory would apply. With one estimate every $\tau_s$ seconds, the number of samples is
$N\simeq T/\tau_s$, and the slope uncertainty scales as
\begin{equation}
\sigma_{\omega_d}^{\rm scat}
\;\simeq\;
C\,
\frac{\sigma_{\phi}^{\rm scat}\sqrt{\tau_s}}{T^{3/2}},
\qquad
\sigma_{\phi}^{\rm scat}\sim \sigma_{P}^{\rm scat}/P_0,
\label{eq:scat_slope_manysamples}
\end{equation}
where $C=\sqrt{12}$ (up to convention) and $P_0$ is the typical polarization magnitude
that sets the mapping between polarization (or asymmetry) and an effective spin angle.

To make the transition from Eq.~(99) to Eq.~(104) mathematically clearer, we can explicitly show the cancellation of the statistical gain. If we attempt to make $N$ independent measurements over time $T$ (where $N = T/\tau_s$), the number of detected particles per sample becomes $N_{\text{det}}/N$. Consequently, the individual error per sample $\sigma_{\phi, i}^{\text{scat}}$ is multiplied by an additional factor of $\sqrt{N}$ compared to Eq.~(99):
\begin{equation}
    \sigma_{\phi, i}^{\text{scat}} \propto \frac{1}{\sqrt{N_{\text{det}}/N}} = \frac{\sqrt{N}}{\sqrt{N_{\text{det}}}} \propto \sqrt{N}
\end{equation}

When we plug this $\sigma_{\phi, i}^{\text{scat}} \propto \sqrt{N}$ into the slope estimation formula:
\begin{equation}
    \sigma_{\omega_d} \propto \frac{\sigma_{\phi, i}^{\text{scat}}}{\sqrt{N} \cdot T}
\end{equation}
the $\sqrt{N}$ terms perfectly cancel out. This proves that regardless of how finely we slice the data, scattering polarimetry mathematically collapses back to:
\begin{equation}
    \sigma_{\omega_d}^{\text{scat}} \propto \frac{1}{T}
\end{equation}
effectively explaining why achieving the $T^{-3/2}$ scaling is fundamentally restricted to non-destructive coherent polarimetry.

In practice, the per-sample uncertainty is large because $N_{\rm det}\ll N_{\rm stored}$
and the analyzing power satisfies $A<1$. For an optimistic illustration, assume a
scattering polarimeter with $100\%$ efficiency and $A=1$. Over a $T=10^3$\,s interval,
suppose the stored beam of $2\times 10^{10}$ protons is extracted uniformly in time with
$\tau_s=1$\,ms spacing, corresponding to $N\simeq 10^6$ samples. Each gate would then
detect $2\times 10^{10}/10^6 \simeq 2\times 10^4$ protons. Using
Eq.~(\ref{eq:scat_slope_manysamples}) with a beam polarization $P_0=0.8$ yields a slope
uncertainty of order $3\times 10^{-8}$\,rad/s ($30$\,nrad/s).

By contrast, a coherent polarimeter can obtain $N\sim 10^6$ independent phase samples
\emph{without} consuming the stored beam, and in the spin-projection-limited regime its
statistical error is smaller by a factor $\sqrt{N}\simeq 10^3$. Extending the spin
coherence time to $T=10^5$\,s would reduce the scattering-based uncertainty to
$\sim 3\times 10^{-10}$\,rad/s, while coherent polarimetry would again provide an
additional improvement of $\sqrt{10^8}\simeq 10^4$ through within-fill slope fitting.

The essential distinction is therefore the scaling with coherent observation time: a
destructive scattering polarimeter effectively achieves only $\sim 1/T$ improvement,
whereas a non-destructive coherent polarimeter realizes the full $T^{-3/2}$ slope
scaling, making SCT prolongation far more powerful.

Including realistic analyzing power ($A<1$) and detection efficiency further increases
the advantage, bringing the total error-reduction factor per-fill to $\sim 1.7\times 10^4$ for
$T=10^3$\,s and $\sim 1.7\times 10^5$ for $T=10^5$\,s, consistent with the estimates
summarized in Table~\ref{tab:SCT}.

\subsubsection{Muon $g\!-\!2$ as a contrasting regime (finite lifetime)}
\label{subsubsec:Statistics_muon_g2}

\paragraph*{Potential muon $g\!-\!2$ experimental sensitivity at 28\,{\rm GeV}.}
A closely analogous situation occurs in the muon $g\!-\!2$ experiments, where the
measurement is intrinsically limited to $\sim 1/T$ scaling because the muons decay,
so that each fill provides effectively only a single polarization-precession estimate
over the lifetime window~\cite{Redin2007NIMA}. In the muon $g\!-\!2$ experiments the spin-precession frequency is extracted from the
time-dependent decay asymmetry, which already constitutes a phase-coherent frequency
estimation over the dilated muon lifetime. A non-destructive coherent polarimeter would
therefore not yield the many-orders-of-magnitude statistical gains possible for long-lived
stored protons, since the coherent observation time remains limited to $T\lesssim\tau_\mu$
($\sim 10^{-3}$~s at tens of GeV). Its primary potential advantage would instead be
systematic: direct spin readout would bypass detector-dependent distortions of the decay
spectrum (pileup, gain drifts, acceptance variations), providing an observable more
closely tied to the instantaneous collective spin phase.

Extending the usable coherent observation window from $T\simeq 64~\mu$s to
$T\simeq 400~\mu$s provides a statistical gain of $\sim 6$ in a one-estimate-per-fill
regime, or up to $\sim 16$ if a true continuous slope estimator can be realized.
With $N_{\rm pol}$ independent coherent pickups distributed around the ring, an
additional $\sqrt{N_{\rm pol}}$ improvement is expected, suggesting a net
$\mathcal{O}(10)$--$\mathcal{O}(50)$ enhancement in statistical reach in an
optimistic multi-polarimeter configuration. These gains are quoted relative to
the conventional scattering-based polarimetry method under otherwise identical
beam conditions and assuming an ideal detector acceptance of $100\%$ for the
conventional approach. For more realistic detector acceptances of order
$\sim10\%$, the advantage of the coherent polarimeter becomes correspondingly
larger.

\subsubsection{The common storage-ring regime: one effective estimate per fill}
\label{subsubsec:Statistics_one_estimate_per_fill}

\paragraph*{The common storage-ring regime: one effective estimate per fill.}
In many practical storage-ring operating modes, scattering polarimetry is intercepting
and therefore depletes the stored beam. As a result, it typically provides at most one
effective polarization (or spin-angle) estimate per fill (or per short gate). In this
regime, the EDM observable is extracted as an accumulated spin-angle (phase) change
over the coherent interval $T$ (see also Appendix~B of~\cite{KimSemertzidis2021PRD}),
\begin{equation}
\Delta\phi \;\simeq\; \omega_d\,T .
\end{equation}
With a per-fill phase uncertainty $\sigma_{\phi}^{\rm scat}$, the corresponding slope
uncertainty is
\begin{equation}
\sigma_{\omega_d}^{\rm scat}
\;\simeq\;
\frac{\sigma_{\phi}^{\rm scat}}{T}
\;\propto\;
\frac{1}{T},
\label{eq:M2_scatt_1overT}
\end{equation}
since there is no additional $1/\sqrt{N}$ statistical gain from repeated,
non-destructive sampling of the \emph{same} stored ensemble within the fill.

\subsubsection{Multiple pickups, SQL, and the meaning of ``independent samples''}
\label{subsubsec:Statistics_SQL_pickups}

\paragraph*{Two-resonator readout and the SQL.}
With two spatially separated resonators interrogating the same stored bunch, the measured
phases may be written as $\phi_{1,2}(t)=\phi_{\rm spin}(t)+n_{1,2}(t)$, where
$\phi_{\rm spin}$ is the common spin-induced phase (including the intrinsic
spin-projection fluctuations of the stored ensemble) and $n_{1,2}$ are the independent
imprecision noises of the two readouts.  Combining or cross-correlating the two resonator
outputs suppresses uncorrelated readout noise (improving the technical floor), but it
does not average down the ensemble SQL itself, because the spin-projection noise is a
property of the beam state and is therefore common to both detectors.  The effective
independent-sampling time is set by the larger of the readout correlation time
$\sim(2B_{\rm meas})^{-1}$ (resonator bandwidth and filtering) and any intrinsic correlation
time of the beam-driven fluctuations.  Using a higher-bandwidth resonator can reduce the
instrumental correlation time, but surpassing the $1/\sqrt{N_{\rm stored}}$ SQL would
require generating sub-Poissonian collective-spin fluctuations (e.g., spin squeezing),
not merely re-measuring the same coherent spin state with multiple polarimeter stations.

The sampling interval $\tau_s$ is set by the measurement bandwidth (resonator response and
demodulation filtering) and defines the spacing of approximately uncorrelated readout
noise samples. The spin-projection SQL sets the minimum achievable phase variance per
sample for the stored ensemble, but it does not correspond to a physical refresh of the
beam on the timescale $\tau_s$.

Successive samples do not represent fresh particle ensembles; rather, they provide a
continuous measurement record whose white phase-noise floor is ultimately bounded by the
spin-projection statistics of the stored beam.

More importantly, coherent polarimetry provides a continuous, non-intercepting phase
time series over the full coherent interval $T$.  In this case the EDM signal is
extracted from a slope fit to a measurement record whose white phase-noise floor is
characterized by a PSD $S_\phi$.  The corresponding Cram\'er--Rao bound yields the
standard slope scaling
\begin{equation}
\sigma_{\omega_d}^{\rm coh}
\;\simeq\;
\sqrt{12}\,\frac{\sqrt{S_\phi}}{T^{3/2}},
\end{equation}
independent of the particular discretization interval used in the analysis.

By contrast, scattering polarimetry is typically intercepting and in many operating
modes provides at most one effective polarization (or spin-angle) estimate per fill.
In that common regime the accumulated angle uncertainty maps onto a slope error
\begin{equation}
\sigma_{\omega_d}^{\rm scat}
\;\simeq\;
\frac{\sigma_{\phi}^{\rm scat}}{T}
\;\propto\;
\frac{1}{T}.
\end{equation}

Thus, beyond the familiar utilization and analyzing-power penalty
($\sigma_{\phi}^{\rm scat}\propto 1/(A\sqrt{\kappa N_{\rm stored}})$),
coherent polarimetry gains additional statistical leverage because it accesses the full
within-fill slope information contained in a broadband measurement record.  The
improvement relative to one-estimate-per-fill scattering may be expressed parametrically
as
\begin{equation}
\frac{\sigma_{\omega_d}^{\rm scat}}{\sigma_{\omega_d}^{\rm coh}}
\;\sim\;
\frac{1}{A\sqrt{\kappa}}
\;\times\;
\sqrt{B\,T},
\label{eq:gain_scat_vs_coh_revised}
\end{equation}
where $B$ is the effective phase-measurement bandwidth set by the coherent readout.
This factor reflects the advantage of continuous non-destructive slope extraction,
not repeated sampling of fresh particle ensembles.
Once technical phase noise is suppressed below the spin-projection floor, coherent
polarimetry can therefore outperform scattering polarimetry by substantially more than
the conventional $\kappa A^2$ considerations alone.

\paragraph{Spin squeezing}
Equation~(\ref{eq:SQL}) assumes an uncorrelated
spin ensemble (SQL). Because the coherent polarimeter is intrinsically non-destructive
and quantum-non-demolition (QND) QND-like with respect to the measured collective spin component, it provides a natural
foundation for future quantum-enhanced protocols (spin squeezing) that can reduce the
projection noise below the SQL~\cite{Wineland1994,Hosten2016}. Such enhancement is not
required for the baseline scaling emphasized here, but becomes relevant once technical
noise sources are sufficiently suppressed and operation approaches the SQL floor.

\subsubsection{Comparison to axion dark matter search sensitivities and estimation of thermal noise sources}
\label{subsubsec:Statistics_axion_thermal_compare}

It is instructive to contrast the phase-noise floor of coherent storage-ring
polarimetry with that achieved in axion dark-matter searches employing resonant
probing.

In axion haloscopes, the dominant noise arises from vacuum and thermal
fluctuations of the cavity electromagnetic field. This noise is essentially
independent of the number of stored particles and, with quantum-limited
amplification and narrowband resonant detection, can be reduced to effective
phase-noise levels in the range
$\sim 10^{-9}$--$10^{-8}$\,rad$/\sqrt{\rm Hz}$~\cite{CARAMEL_PRD2026}.
In the CARAMEL framework, the use of a coherent RF probe together with optical
electro-optic readout improves the effective signal-to-noise by a factor of
$\sim 5$--$10$ relative to conventional microwave-amplifier chains, corresponding
to a $\sim 25$--$100\times$ enhancement in scanning speed at fixed target
sensitivity (since scan rate scales as ${\rm SNR}^2$)~\cite{CARAMEL_PRD2026}.

By contrast, in coherent polarimetry the measured phase is a collective spin
observable. The fundamental noise floor is therefore set by spin-projection
noise of the stored ensemble, scaling as
$1/(\sqrt{N_{\rm stored}})$. For representative storage-ring parameters
($P\simeq 0.8$ and $N_{\rm stored}\simeq 2\times 10^{10}$), the corresponding
per-sample phase uncertainty is given in Eq.~\ref{eq:SQL_inst}.
Expressed as an equivalent white phase-noise density for sampling at the
resonator-limited rate $\tau_s\simeq 1$\,ms (bandwidth $B\simeq (2\tau_s)^{-1}$),
this corresponds to
\begin{equation}
\sqrt{S_\phi^{\rm(SQL)}} \;\simeq\; \sigma_\phi^{\rm(SQL)}\sqrt{2\tau_s}
\;\approx\; 4\times 10^{-7}\ \mathrm{rad}/\sqrt{\mathrm{Hz}}.
\label{eq:SQL_perHz}
\end{equation}

The difference reflects the nature of the quantum object being interrogated:
axion experiments measure a classical electromagnetic field in the presence of
vacuum fluctuations, whereas coherent polarimetry measures the quantum state of
a macroscopic spin ensemble. In both cases, resonant probing enables
phase-sensitive readout near the quantum limit, but the ultimate noise floor is
set by fundamentally different degrees of freedom.

\paragraph{Thermal noise of the pickup electrodes}
A useful quantitative point is that room-temperature operation of the pickup
electrodes does not automatically preclude reaching the spin-projection limit.
At $f_0\simeq 18.18$~MHz, the thermal photon occupation number at $T=300$~K is
\begin{equation}
\bar n_{\rm th}
=\frac{1}{e^{\hbar\omega/k_BT}-1}
\simeq \frac{k_BT}{\hbar\omega}
\approx 3\times 10^{5},
\end{equation}
so the cavity electromagnetic field is deeply in the classical (thermal) regime.
The corresponding phase fluctuations of a room-temperature resonator would be
enormous \emph{if} one were attempting to measure the cavity vacuum field itself,
as in an axion haloscope.
This is precisely why axion haloscopes typically require cryogenic operation:
in that context, thermal photons constitute the dominant irreducible noise
source of the electromagnetic mode.

In coherent polarimetry we drive the resonator with a strong carrier and measure
the small phase modulation imprinted by the polarized beam.  We therefore do
not attempt to resolve the cavity’s intrinsic thermal field amplitude; thermal
and Johnson noise only matter to the extent that they add phase noise to the
carrier near the spin-wheel band, and become relevant only if they exceed the
spin-projection floor of Eq.~\ref{eq:SQL_perHz}.

A more explicit way to quantify the room-temperature requirement is to estimate
the Johnson--Nyquist noise of the pickup electrodes and compare it directly to
the spin-projection phase-noise floor.

At $T=300$~K, a resistive pickup with impedance $R\simeq 50~\Omega$ produces a
(one-sided) thermal voltage noise spectral density
\begin{equation}
\sqrt{S_V^{\rm th}}=\sqrt{4k_BT R}
\simeq 9\times 10^{-10}\ \mathrm{V}/\sqrt{\mathrm{Hz}} .
\end{equation}
In a phase-sensitive measurement, the relevant quantity is the corresponding
phase noise on the driven carrier. For a carrier voltage amplitude $V_c$, small
additive voltage noise produces an effective phase noise density of order
\begin{equation}
\sqrt{S_\phi^{\rm th}}
\;\sim\;
\frac{\sqrt{S_V^{\rm th}}}{V_c}.
\end{equation}

For illustration, consider a modest carrier power of $P_c=1$~mW (0~dBm)
delivered into $50~\Omega$. The corresponding rms carrier voltage is
\begin{equation}
V_c=\sqrt{P_c R}\simeq 0.22\ \mathrm{V},
\end{equation}
which gives
\begin{equation}
\sqrt{S_\phi^{\rm th}(300K)}
\sim
\frac{9\times 10^{-10}}{0.22}
\approx 4\times 10^{-9}\ \mathrm{rad}/\sqrt{\mathrm{Hz}}.
\label{eq:thermal_noise}
\end{equation}
This is already two orders of magnitude \emph{below} the spin-projection-limited
phase-noise density for the stored ensemble, given in Eq.~\ref{eq:SQL_perHz}.
Thus, with even milliwatt-level carrier power, room-temperature Johnson noise of
the pickup electrodes is not the limiting factor: the measurement remains
spin-statistics dominated.

This comparison also indicates how much spin squeezing could be exploited before
room-temperature thermal noise becomes relevant. A squeezing gain $G_{\rm sq}$
reduces the spin-projection floor by $\sqrt{G_{\rm sq}}$, so the crossover occurs
when
\begin{equation}
\frac{\sqrt{S_\phi^{\rm(SQL)}}}{\sqrt{G_{\rm sq}}}
\sim
\sqrt{S_\phi^{\rm th}(300K)}.
\end{equation}
Using the values above yields
\begin{equation}
G_{\rm sq}^{\rm(max)}
\sim
\left(\frac{4\times 10^{-7}}{4\times 10^{-9}}\right)^2
\sim 10^{4}.
\end{equation}
Therefore, one could in principle achieve up to $\sim 40$~dB of spin squeezing
before the intrinsic Johnson noise of a room-temperature $50~\Omega$ pickup
electrode would begin to compete with the quantum spin-projection floor. In
practice, other technical noise sources (amplifier noise, oscillator phase
noise, microphonics) will enter well before this level, but the estimate shows
that cryogenic operation of the pickup electrodes is not a fundamental
requirement for projection-noise-limited coherent polarimetry.

Unlike axion haloscopes, coherent polarimetry does not require cryogenic operation to suppress thermal photon noise, since the observable is a phase modulation on a driven carrier rather than the cavity’s intrinsic field amplitude. Cryogenic or HTS resonators are therefore motivated primarily by stability and achievable quality factor $Q$, not by a fundamental thermal-photon requirement.
It is important to emphasize that thermal and vacuum fluctuations of the cavity
field are present in both axion haloscopes and coherent polarimetry. The key
distinction is the manner in which the signal enters the driven resonator. In
axion searches the signal appears as an \emph{additive} cavity field sourced by
the axion effective current, and therefore competes directly with the thermal
photon occupation of the electromagnetic mode. In coherent polarimetry, by
contrast, the polarized beam produces primarily a \emph{phase modulation} of an
already large driven carrier, so that the relevant noise is the fractional
phase imprecision of the probe. With sufficient carrier amplitude, thermal
field fluctuations contribute only a subdominant technical phase noise and need
not require cryogenic suppression in the same way as in passive axion power
detection.

\paragraph{Two-resonator readout and the SQL.}
With two spatially separated resonant pickups interrogating the same stored ensemble, the
measured phases may be written as $\phi_{1,2}(t)=\phi_{\rm spin}(t)+n_{1,2}(t)$, where
$\phi_{\rm spin}$ is the common spin-induced phase (including intrinsic spin-projection
fluctuations of the stored ensemble) and $n_{1,2}$ are independent imprecision noises of
the two readouts. Combining or cross-correlating the two outputs suppresses uncorrelated
readout noise (improving the technical floor), but it does not average down the ensemble
SQL itself, because the spin-projection noise is a property of the beam state and is
therefore common to both detectors. Surpassing the $1/\sqrt{N_{\rm stored}}$ SQL would
require generating sub-Poissonian collective-spin fluctuations (e.g., spin squeezing),
not merely re-measuring the same coherent spin state with multiple pickups~\cite{Wineland1994,Hosten2016}.

\subsection{Spin wheel, intra-beam scattering, and stochastic cooling}
\label{subsec:SW_IBS_cooling}

A controlled vertical spin precession (``spin wheel'', SW) at a frequency in the
range of order $0.1$--$10\,\mathrm{Hz}$ is introduced to provide a stable phase
reference for synchronous detection and to shift the measurement away from DC.
The spin wheel does not amplify the EDM signal itself, which appears as a
\emph{slope} in the phase evolution of the spin vector.
Instead, the controlled precession establishes the phase reference against
which the EDM-induced phase drift is measured.
Operating at a finite spin frequency enables phase-referenced detection and
provides powerful systematic checks through controlled variation of the SW
frequency and reversal of the rotation direction.

Intra-beam scattering (IBS) drives diffusion of orbital phase-space variables,
including the momentum spread and the transverse emittances, on characteristic
timescales that depend strongly on beam energy, intensity, and lattice design.
At the relatively low beam energies considered for compact storage-ring EDM
concepts, IBS is generally strong: growth times of order
$\sim 1$--$10\,\mathrm{s}$ are typical in unoptimized lattices, with rates
increasing rapidly toward lower energies (approximately scaling as
$\beta^{-2}$).
Consequently, careful lattice design together with active cooling is essential
in order to extend these timescales to values compatible with long
coherence-based measurements.

For the symmetric hybrid ring lattice optimized specifically for EDM
operation, detailed studies indicate that IBS growth times can reach
$\mathcal{O}(10^{3}\,\mathrm{s})$ under realistic operating conditions
\cite{Omarov_2022_symmetric}.
Such optimization is an essential ingredient for enabling long storage times,
although the detailed impact of IBS-driven orbital diffusion on the spin
dynamics remains an active topic that requires comprehensive spin--orbit
tracking simulations.

Because the spin-wheel period lies in the range
$\sim 10^{-1}$--$10\,\mathrm{s}$ for typical operating frequencies,
IBS-induced variations occur slowly compared to the controlled spin motion and
therefore act primarily as quasi-static perturbations over many spin cycles.
Stochastic cooling acts directly on the relevant orbital degrees of freedom,
reducing the growth of momentum spread and transverse amplitudes and thereby
stabilizing the stored beam over long durations.

In this framework, the combination of controlled spin precession, active beam
cooling, and coherence-preserving control sequences is expected to maintain
spin phase coherence over macroscopic timescales.
In favorable scenarios the effective spin coherence time (SCT) may approach
$10^{4}$--$10^{5}\,\mathrm{s}$.
The role of the coherent polarimeter is then to provide a readout that remains
linear and phase-stable over the same interval, allowing the phase-slope
estimator to exploit the full coherent integration time
\cite{Schmitt2017T32}.

The conceptual advantage of introducing a controlled spin precession was
recognized early on.
By shifting the measurement away from DC and enabling phase-referenced
detection, the effective coherence time can be extended in a natural way
\cite{Koop:2013vja,Koop:2015lez}.
This insight remains a cornerstone of storage-ring EDM experiments.

Early discussions of the spin-wheel concept often considered operation at
relatively low beam energies, motivated by compact ring designs.
In such regimes, however, strong IBS at low $\beta$ leads to rapid growth of
transverse and longitudinal emittances and correspondingly short beam
lifetimes.
Moreover, away from the magic momentum, orbital diffusion couples more
directly into spin-tune spread and spin-phase decoherence.

The present strategy combines spin-wheel operation with an IBS-optimized
lattice, active stochastic cooling, and coherence-preserving control
techniques.
Spin-echo methods, widely used in atomic clocks and precision spectroscopy,
can refocus ensemble dephasing arising from slowly varying frequency shifts
and field inhomogeneities \cite{Hahn1950,LudlowRevModPhys2015}.
When the dominant phase evolution is \emph{reversible}, echo sequences provide
a powerful mechanism for recovering coherence.

It is important to emphasize that spin echo does not eliminate truly stochastic
spin diffusion in the strict sense of a random walk: rapidly fluctuating,
uncorrelated spin-phase noise cannot be time-reversed.
Instead, echo sequences mitigate quasi-static and slowly varying components of
the spin-tune spread---including those arising indirectly from IBS-driven
orbital evolution---while the residual fast diffusion ultimately determines
the fundamental coherence limit.
Quantifying this interplay in realistic lattices remains an important subject
for detailed simulation and experimental validation.

These developments build directly on the original spin-wheel concept while
incorporating advances in lattice optimization, beam cooling, and
coherence-preserving techniques that are essential for exploiting long
coherent integration times in next-generation storage-ring EDM searches.

\subsection{Coherent polarimeter readout strategy}
\label{subsec:coherent_readout}

The coherent polarimeter is operated explicitly as a phase-sensitive
measurement.
A stable probe tone establishes a resonant carrier at the resonator frequency
$f_0$, defining a long-lived phase reference against which small
spin-dependent perturbations can be measured.
The polarized beam induces a helicity-dependent modification of the
electromagnetic boundary condition at the pickup electrodes, which appears as a
small, correlated phase perturbation of the driven resonator field.

The readout proceeds through synchronous demodulation referenced to the known
temporal structure of the spin dynamics.
This includes the spin-wheel phase as well as the timing markers associated
with CW/CCW beam circulation and helicity reversal.
By projecting the data onto this known temporal structure, the analysis
selectively retains components that are phase-coherent with the spin motion
while rejecting signals that are not correlated with it.

Information is accumulated coherently in a phase-slope (frequency) estimator
over the full spin coherence interval.
Because the EDM signal appears as a small change in the phase evolution rate,
the sensitivity improves with observation time according to the
$T^{-3/2}$ scaling discussed previously.
This coherent accumulation is enabled by the stability of the probing carrier
and by the narrowband filtering provided by the resonator.

In this architecture the resonator functions primarily as a linear phase
transducer and a stable reference element rather than as an energy-storage
device for a beam-driven signal.
The probing field continuously interrogates the boundary condition imposed by
the polarized beam, converting the spin-dependent electromagnetic perturbation
into a measurable phase modulation of the carrier.
The EDM observable is therefore extracted from the accumulated phase evolution
of a collective electromagnetic mode, enabling continuous, non-intercepting
polarimetry that remains compatible with long spin-coherence times.

\bibliography{references}

@article{Kimball2016_PhysRevLett.116.190801,
  title = {Precessing Ferromagnetic Needle Magnetometer},
  author = {Jackson Kimball, Derek F. and Sushkov, Alexander O. and Budker, Dmitry},
  journal = {Phys. Rev. Lett.},
  volume = {116},
  issue = {19},
  pages = {190801},
  numpages = {7},
  year = {2016},
  month = {May},
  publisher = {American Physical Society},
  doi = {10.1103/PhysRevLett.116.190801},
  url = {https://link.aps.org/doi/10.1103/PhysRevLett.116.190801}
}

@article{Ni2025_1v1p-kpb2,
  title = {Microscopic theory of a precessing ferromagnet for ultrasensitive magnetometry},
  author = {Ni, Xueqi and Zou, Zhixing and Lecamwasam, Ruvi and Vinante, Andrea and Budker, Dmitry and Lam, Ping Koy and Wang, Tao and Gong, Jiangbin},
  journal = {Phys. Rev. Res.},
  volume = {7},
  issue = {4},
  pages = {043120},
  numpages = {12},
  year = {2025},
  month = {Oct},
  publisher = {American Physical Society},
  doi = {10.1103/1v1p-kpb2},
  url = {https://link.aps.org/doi/10.1103/1v1p-kpb2}
}

@article{Redin2007NIMA,
  title        = {Statistical equations and methods applied to the precision muon $(g-2)$ experiment at {BNL}},
  author       = {Bennett, G. W. and others},
  journal      = {Nuclear Instruments and Methods in Physics Research Section A},
  volume       = {579},
  pages        = {1096--1116},
  year         = {2007},
  doi          = {10.1016/j.nima.2007.06.023}
}

@article{KimSemertzidis2021PRD,
  title        = {New method of probing an oscillating {EDM} induced by axionlike dark matter using an rf {Wien} filter in storage rings},
  author       = {Kim, On and Semertzidis, Yannis K.},
  journal      = {Physical Review D},
  volume       = {104},
  pages        = {096006},
  year         = {2021},
  doi          = {10.1103/PhysRevD.104.096006}
}

@article{ProbingPRD2023,
  title        = {Speeding axion haloscope experiments using heterodyne-variance-based detection with a power meter},
  author       = {Omarov, Zhanibek and others},
  journal      = {Physical Review D},
  volume       = {107},
  pages        = {103005},
  year         = {2023},
  doi          = {10.1103/PhysRevD.107.103005}
}

@article{CARAMEL_PRD2026,
  title = {Cosmic axions revealed via amplified modulation of the ellipticity of a laser},
  author = {Davoudiasl, Hooman and Semertzidis, Yannis K.},
  journal = {Phys. Rev. D},
  volume = {113},
  issue = {3},
  pages = {032012},
  numpages = {18},
  year = {2026},
  month = {Feb},
  publisher = {American Physical Society},
  doi = {10.1103/2dvg-78kh},
  url = {https://link.aps.org/doi/10.1103/2dvg-78kh}
}

@misc{CARAMEL2025Preprint,
  title        = {Cosmic Axions Revealed via Amplified Modulation of Ellipticity of Laser (CARAMEL)},
  author       = {Hooman Davoudiasl and Semertzidis, Yannis K.},
  year         = {2025},
      eprint={2506.24022},
      archivePrefix={arXiv},
      primaryClass={hep-ph},
url     = {https://doi.org/10.48550/arXiv.2506.24022}
}

@article{Omarov_2022_symmetric,
  title = {Comprehensive symmetric-hybrid ring design for a proton EDM experiment at below ${10}^{\ensuremath{-}29}\mathrm{e}\ifmmode\cdot\else\textperiodcentered\fi{}\mathrm{cm}$},
  author = {Omarov, Zhanibek and Davoudiasl, Hooman and Hac\ifmmode \imath \else \i \fi{}\"omero\ifmmode \breve{g}\else \u{g}\fi{}lu, Selcuk and Lebedev, Valeri and Morse, William M. and Semertzidis, Yannis K. and Silenko, Alexander J. and Stephenson, Edward J. and Suleiman, Riad},
  journal = {Phys. Rev. D},
  volume = {105},
  issue = {3},
  pages = {032001},
  numpages = {20},
  year = {2022},
  month = {Feb},
  publisher = {American Physical Society},
  doi = {10.1103/PhysRevD.105.032001},
  url = {https://link.aps.org/doi/10.1103/PhysRevD.105.032001}
}

@article{Karananas2025,
  author = {Karananas, Georgios K. and Shaposhnikov, Mikhail and Zell, Sebastian},
  title = {Gravitational {O}rigin of the {QCD} {A}xion},
  journal = {Phys. Rev. Lett.},
  volume = {135},
  pages = {241001},
  year = {2025},
  doi = {10.1103/PhysRevLett.135.241001},
  eprint = {2507.04662},
  archivePrefix = {arXiv},
  primaryClass = {hep-ph}
}

@article{Karananas2025_noncompact,
  author = {Karananas, Georgios K. and Shaposhnikov, Mikhail and Zell, Sebastian},
  title = {A non-compact {QCD} axion},
  journal = {arXiv preprint arXiv:2512.20290},
  year = {2025},
  eprint = {2512.20290},
  archivePrefix = {arXiv},
  primaryClass = {hep-ph}
}

@article{Wineland1994,
  title = {Squeezed atomic states and projection noise in spectroscopy},
  author = {Wineland, D. J. and Bollinger, J. J. and Itano, W. M. and Heinzen, D. J.},
  journal = {Phys. Rev. A},
  volume = {50},
  issue = {1},
  pages = {67--88},
  numpages = {0},
  year = {1994},
  month = {Jul},
  publisher = {American Physical Society},
  doi = {10.1103/PhysRevA.50.67},
  url = {https://link.aps.org/doi/10.1103/PhysRevA.50.67}
}

@article{Hosten2016,
  title        = {Measurement noise 100 times lower than the quantum-projection limit using entangled atoms},
  author       = {Hosten, O. and others},
  journal      = {Nature},
  volume       = {529},
  pages        = {505--508},
  year         = {2016},
  doi          = {10.1038/nature16176}
}

@article{Hahn1950,
  author = {E. L. Hahn},
  title = {Spin Echoes},
  journal = {Phys. Rev.},
  volume = {80},
  page = {580},
  year = {1950}
}

@article{JEDI:2016swi,
    author = "Guidoboni, G. and others",
    collaboration = "JEDI",
    title = "{How to Reach a Thousand-Second in-Plane Polarization Lifetime with 0.97-GeV/c Deuterons in a Storage Ring}",
    doi = "10.1103/PhysRevLett.117.054801",
    journal = "Phys. Rev. Lett.",
    volume = "117",
    pages = "054801",
    year = "2016"
}

@inproceedings{derbenev1998radio,
  title={Radio-frequency polarimetry},
  author={Derbenev, Ya S},
  booktitle={AIP Conference Proceedings},
  volume={421},
  pages={191--199},
  year={1998},
  organization={American Institute of Physics}
}

@techreport{cameron1996absolute,
  title={Absolute Calibration and Beam Background of the Squid Polarimeter},
  author={Cameron, PR and Blaskiewicz, MM and Derbenev, Ya S and Goldberg, DA and Luccio, AU and Mariam, FG and Shea, TJ and Syphers, MJ and Tsoupas, N},
  year={1996},
  institution={Brookhaven National Laboratory (BNL), Upton, NY (United States)}
}

@inproceedings{cameron1997,
  title={Squids, snakes and    polarimeters: a new Technique for measuring the magnetic moments of polarized beams},
  author={Cameron, PR and Luccio, AU and Shea, TJ and Tsoupas, N and Goldberg, DA},
  booktitle={AIP Conference Proceedings},
  volume={390},
  pages={306--315},
  year={1997},
  organization={American Institute of Physics}
}

@techreport{blaskiewicz1996absolute,
  title={Absolute calibration and beam background of the Squid Polarimeter},
  author={Blaskiewicz, MM and Cameron, PR and Shea, TJ},
  year={1996},
  institution={Brookhaven National Lab.(BNL), Upton, NY (United States); Lawrence Berkeley~…}
}

@techreport{Roberts2021Polarimeter,
  author = {B. Roberts},
  title = {Resonant Magnetometry and Polarimetry},
  institution = {DOE Nuclear Physics SBIR},
  number = {DE-SC0017120},
  year = {2021}
}

@article{Koop:2015lez,
    author = "Koop, I. A.",
    editor = "Egelhof, Peter and Litvinov, Yuri and Steck, Markus",
    title = "{Colliding or co-rotating ion beams in storage rings for EDM search}",
    doi = "10.1088/0031-8949/2015/T166/014034",
    journal = "Phys. Scripta T",
    volume = "166",
    pages = "014034",
    year = "2015"
}

@inproceedings{Koop:2013vja,
    author = "Koop, Ivan",
    title = "{Asymmetric Energy Colliding Ion Beams in the EDM Storage Ring}",
    booktitle = "{4th International Particle Accelerator Conference}",
    pages = "TUPWO040",
    year = "2013"
}

@article{Farley_PhysRevLett.93.052001,
  title = {New Method of Measuring Electric Dipole Moments in Storage Rings},
  author = {Farley, F. J. M. and Jungmann, K. and Miller, J. P. and Morse, W. M. and Orlov, Y. F. and Roberts, B. L. and Semertzidis, Y. K. and Silenko, A. and Stephenson, E. J.},
  journal = {Phys. Rev. Lett.},
  volume = {93},
  issue = {5},
  pages = {052001},
  numpages = {4},
  year = {2004},
  month = {Jul},
  publisher = {American Physical Society},
  doi = {10.1103/PhysRevLett.93.052001},
  url = {https://link.aps.org/doi/10.1103/PhysRevLett.93.052001}
}

@article{LudlowRevModPhys2015,
  title = {Optical atomic clocks},
  author = {Ludlow, Andrew D. and Boyd, Martin M. and Ye, Jun and Peik, E. and Schmidt, P. O.},
  journal = {Rev. Mod. Phys.},
  volume = {87},
  issue = {2},
  pages = {637--701},
  numpages = {65},
  year = {2015},
  month = {Jun},
  publisher = {American Physical Society},
  doi = {10.1103/RevModPhys.87.637},
  url = {https://link.aps.org/doi/10.1103/RevModPhys.87.637}
}

@article{RifeBoorstyn1974,
  author  = {Rife, D. C. and Boorstyn, R. R.},
  title   = {Single tone parameter estimation from discrete-time observations},
  journal = {IEEE Transactions on Information Theory},
  volume  = {20},
  pages   = {591--598},
  year    = {1974}
}

@article{Quinn1994,
  author  = {Barry G. Quinn},
  title   = {Threshold behavior of the maximum likelihood estimator of frequency},
  journal = {IEEE Transactions on Signal Processing},
  volume  = {42},
  number  = {11},
  pages   = {3291--3294},
  year    = {1994},
  doi     = {10.1109/78.330368}
}

@article{Tretter1985,
  author  = {Steven A. Tretter},
  title   = {Estimating the frequency of a noisy sinusoid by linear regression},
  journal = {IEEE Transactions on Information Theory},
  volume  = {31},
  number  = {6},
  pages   = {832--835},
  year    = {1985},
  doi     = {10.1109/TIT.1985.1057115}
}

@article{Boashash1992,
  author  = {Boualem Boashash},
  title   = {Estimating and interpreting the instantaneous frequency of a signal. {I}. Fundamentals},
  journal = {Proceedings of the IEEE},
  volume  = {80},
  number  = {4},
  pages   = {520--538},
  year    = {1992},
  doi     = {10.1109/5.135376}
}

@article{Schmitt2017T32,
author = {Simon Schmitt  and Tuvia Gefen  and Felix M. Stürner  and Thomas Unden  and Gerhard Wolff  and Christoph Müller  and Jochen Scheuer  and Boris Naydenov  and Matthew Markham  and Sebastien Pezzagna  and Jan Meijer  and Ilai Schwarz  and Martin Plenio  and Alex Retzker  and Liam P. McGuinness  and Fedor Jelezko },
title = {Submillihertz magnetic spectroscopy performed with a nanoscale quantum sensor},
journal = {Science},
volume = {356},
number = {6340},
pages = {832-837},
year = {2017},
doi = {10.1126/science.aam5532},
URL = {https://www.science.org/doi/abs/10.1126/science.aam5532},
eprint = {https://www.science.org/doi/pdf/10.1126/science.aam5532}}

@book{Kay1993,
  author = {S. M. Kay},
  title = {Fundamentals of Statistical Signal Processing, Volume I: Estimation Theory},
  publisher = {Prentice Hall},
  year = {1993}
}

@article{Jeong:2022,
  author  = {H. Jeong and E. Won and Y. K. Semertzidis and E. J. Stephenson and S. Park},
  title   = {An implementation of spin-dependent hadron elastic scattering in {GEANT4}},
  journal = {J. Korean Phys. Soc.},
  volume  = {80},
  pages   = {437},
  year    = {2022},
  doi     = {10.1007/s40042-022-00429-7}
}

@article{Graham:2013_PhysRevD.88.035023,
  title = {New observables for direct detection of axion dark matter},
  author = {Graham, Peter W. and Rajendran, Surjeet},
  journal = {Phys. Rev. D},
  volume = {88},
  issue = {3},
  pages = {035023},
  numpages = {13},
  year = {2013},
  month = {Aug},
  publisher = {American Physical Society},
  doi = {10.1103/PhysRevD.88.035023},
  url = {https://link.aps.org/doi/10.1103/PhysRevD.88.035023}
}

@article{Graham:2021_PhysRevD.103.055010,
  title = {Storage ring probes of dark matter and dark energy},
  author = {Graham, Peter W. and Hac\ifmmode \imath \else \i \fi{}\"omero\ifmmode \breve{g}\else \u{g}\fi{}lu, Selcuk and Kaplan, David E. and Omarov, Zhanibek and Rajendran, Surjeet and Semertzidis, Yannis K.},
  journal = {Phys. Rev. D},
  volume = {103},
  issue = {5},
  pages = {055010},
  numpages = {13},
  year = {2021},
  month = {Mar},
  publisher = {American Physical Society},
  doi = {10.1103/PhysRevD.103.055010},
  url = {https://link.aps.org/doi/10.1103/PhysRevD.103.055010}
}

@article{Allan1966,
  author = {Allan, David W.},
  title = {Statistics of Atomic Frequency Standards},
  journal = {Proceedings of the IEEE},
  volume = {54},
  number = {2},
  pages = {221--230},
  year = {1966},
  doi = {10.1109/PROC.1966.4634}
}

@article{Gerberding2015,
  author = {Gerberding, Oliver and Diekmann, Christoph and Bykov, Igor and Delgado, Juan Jose Esteban and Steger, Martin and others},
  title = {Readout for intersatellite laser interferometry: Measuring low frequency phase fluctuations of HF signals with microradian precision},
  journal = {Physical Review D},
  volume = {91},
  pages = {042003},
  year = {2015},
  doi = {10.1103/PhysRevD.91.042003}
}

@article{Gerberding2013,
  author = {Gerberding, Oliver and Sheard, Benjamin and Heinzel, Gerhard},
  title = {A phasemeter for laser heterodyne interferometry in space},
  journal = {Review of Scientific Instruments},
  volume = {84},
  pages = {074501},
  year = {2013},
  doi = {10.1063/1.4813805}
}

@article{Tinto2014,
  author = {Tinto, Massimo and Dhurandhar, Sanjeev V.},
  title = {Time-delay interferometry},
  journal = {Living Reviews in Relativity},
  volume = {17},
  pages = {6},
  year = {2014},
  doi = {10.12942/lrr-2014-6}
}

@article{Huang_harmonic_PhysRevAccelBeams.19.084201,
  title = {Ultrafast harmonic rf kicker design and beam dynamics analysis for an energy recovery linac based electron circulator cooler ring},
  author = {Huang, Yulu and Wang, Haipeng and Rimmer, Robert A. and Wang, Shaoheng and Guo, Jiquan},
  journal = {Phys. Rev. Accel. Beams},
  volume = {19},
  issue = {8},
  pages = {084201},
  numpages = {10},
  year = {2016},
  month = {Aug},
  publisher = {American Physical Society},
  doi = {10.1103/PhysRevAccelBeams.19.084201},
  url = {https://link.aps.org/doi/10.1103/PhysRevAccelBeams.19.084201}
}

@techreport{Bovet:1970,
  author       = {C. Bovet and R. Gouiran and I. Gumowski and K.H. Reich},
  title        = {A Selection of Formulae and Data Useful for the Design of A.G. Synchrotrons},
  institution  = {CERN},
  address      = {Geneva},
  year         = {1970},
  number       = {CERN-MPS-SI-Int-DL-70-4},
  url          = {https://cds.cern.ch/record/104153/files/MPS-SI-Int-DL-70-4.pdf}
}

@article{PTOLEMY_HV2026,
doi = {10.1088/1748-0221/21/04/P04009},
url = {https://doi.org/10.1088/1748-0221/21/04/P04009},
year = {2026},
month = {apr},
publisher = {IOP Publishing},
volume = {21},
number = {04},
pages = {P04009},
author = {Ammendola, R. and Apponi, A. and Benato, G. and Betti, M.G. and Biondi, R. and Bos, P. and Cavoto, G. and Cadeddu, M. and Casale, A. and Castellano, O. and Celasco, E. and Cecchini, L. and Chirico, M. and Chung, W. and Cocco, A.G. and Colijn, A.P. and Corcione, B. and D'Ambrosio, N. and D'Incecco, M. and De Bellis, G. and De Deo, M. and de Groot, N. and Esposito, A. and Farino, M. and Farinon, S. and Ferella, A.D. and Ferro, L. and Ficcadenti, L. and Galbato Muscio, G. and Gariazzo, S. and Garrone, H. and Gatti, F. and Korga, G. and Malnati, F. and Mangano, G. and Marcucci, L.E. and Mariani, C. and Mead, J. and Menichetti, G. and Messina, M. and Monticone, E. and Naafs, M. and Narcisi, V. and Nagorny, S. and Neri, G. and Pandolfi, F. and Pavarani, R. and Pérez de los Heros, C. and Pisanti, O. and Pepe, C. and Pofi, F.M. and Polosa, A.D. and Rago, I. and Rajteri, M. and Rossi, N. and Ritarossi, S. and Ruocco, A. and Salina, G. and Santucci, A. and Sestu, M. and Tan, A. and Tozzini, V. and Tully, C.G. and van Rens, I. and Virzi, F. and Visser, G. and Viviani, M.},
collaboration = {The PTOLEMY Collaboration},
title = {Ultra-high precision high voltage system for {PTOLEMY}},
journal = {Journal of Instrumentation}
}

@article{PLL_ADI,
  author = {Ian Collins},
  title = {{Phase-Locked Loop} ({PLL}) {Fundamentals}},
  journal = {Analog Dialogue},
  year = {2018},
  url = {https://www.analog.com/en/resources/analog-dialogue/articles/phase-locked-loop-pll-fundamentals.html}
}

\end{document}